\documentclass[12pt]{article}
\usepackage[a4paper, hmargin=2.5cm, vmargin=2.5cm]{geometry}
\usepackage{graphicx,amssymb,amstext,amsmath}
\usepackage{url, comment, caption}
\usepackage{amsfonts}
\usepackage[authoryear]{natbib} \bibpunct{(}{)}{,}{a}{,}{,}
\usepackage{makecell}

\newcommand{\ba}{\mbox{\boldmath $a$}}
\newcommand{\bc}{\mbox{\boldmath $c$}}
\newcommand{\bd}{\mbox{\boldmath $d$}}
\newcommand{\be}{\mbox{\boldmath $e$}}
\newcommand{\bu}{\mbox{\boldmath $u$}}
\newcommand{\bw}{\mbox{\boldmath $w$}}
\newcommand{\bx}{\mbox{\boldmath $x$}}
\newcommand{\by}{\mbox{\boldmath $y$}}
\newcommand{\bz}{\mbox{\boldmath $z$}}
\newcommand{\bB}{\mbox{\boldmath $B$}}
\newcommand{\bC}{\mbox{\boldmath $C$}}
\newcommand{\bD}{\mbox{\boldmath $D$}}
\newcommand{\bI}{\mbox{\boldmath $I$}}
\newcommand{\bJ}{\mbox{\boldmath $J$}}
\newcommand{\bU}{\mbox{\boldmath $U$}}
\newcommand{\bX}{\mbox{\boldmath $X$}}
\newcommand{\bY}{\mbox{\boldmath $Y$}}
\newcommand{\bZ}{\mbox{\boldmath $Z$}}
\newcommand{\bPsi}{\mbox{\boldmath $\Psi$}}
\newcommand{\btheta}{\mbox{\boldmath $\theta$}}
\newcommand{\bepsilon}{\mbox{\boldmath $\epsilon$}}
\newcommand{\bmu}{\mbox{\boldmath $\mu$}}
\newcommand{\bSigma}{\mbox{\boldmath $\Sigma$}}
\newcommand{\bOmega}{\mbox{\boldmath $\Omega$}}

\newcommand{\bDelta}{\mbox{\boldmath $\Delta$}}
\newcommand{\bLambda}{\mbox{\boldmath $\Lambda$}}
\newcommand{\bzeta}{\mbox{\boldmath $\zeta$}}
\newcommand{\bzero}{\mbox{\boldmath $0$}}
\newcommand{\bOne}{\mbox{\boldmath $1$}}
\newcommand{\bpi}{\mbox{\boldmath $\pi$}}
\newcommand{\bpsi}{\mbox{\boldmath $\psi$}}
\newcommand{\hbPsi}{\hat{\mbox{\boldmath $\bPsi$}}}

\begin{document}
\title{Mixtures of Factor Analyzers with Fundamental Skew Symmetric Distributions}
\author{Sharon X. Lee$^1$, Tsung-I Lin$^{2,3}$, Geoffrey J. McLachlan$^{1, \star}$}
\date{}

\maketitle

\begin{flushleft}
$^1$Department of Mathematics,
University of Queensland, St. Lucia, Queensland, 4072, Australia.\\
$^2$Institute of Statistics,
National Chung Hsing University, Taiwan.\\
$^3$Department of Public Health, China Medical University, Taichung, Taiwan. \\
$^\star$ E-mail: g.mclachlan@uq.edu.au 
\end{flushleft}

\begin{abstract}
Mixtures of factor analyzers (MFA) provide a powerful tool 
for modelling high-dimensional datasets. 
In recent years, several generalizations of MFA 
have been developed where the normality assumption of 
the factors and/or of the errors was relaxed to allow for 
skewness in the data. However, due to the form of 
the adopted component densities, 
the distribution of the factors/errors in most of these models 
is typically limited to modelling skewness 
concentrated in a single direction.  
Here, we introduce a more flexible finite mixture of factor analyzers 
based on the class of scale mixtures of 
canonical fundamental skew normal (SMCFUSN) distributions. 
This very general class of skew distributions 
can capture various types of skewness and asymmetry in the data. 
In particular, the proposed mixture model of SMCFUSN factor analyzers 
(SMCFUSNFA) can simultaneously 
accommodate multiple directions of skewness. 
As such, it encapsulates many commonly used models 
as special and/or limiting cases, 
such as models of some versions of skew normal 
and skew $t$-factor analyzers, 
and skew hyperbolic factor analyzers. 
For illustration, we focus on 
the $t$-distribution member of 
the class of SMCFUSN distributions, 
leading to mixtures of canonical fundamental skew $t$-factor analyzers (CFUSTFA).
Parameter estimation can be carried out 
by maximum likelihood via an EM-type algorithm. 
The usefulness and potential of the proposed model 
are demonstrated using  
two real datasets. 
\end{abstract}

\section{\large Introduction}
\label{s:intro}

The factor analysis (FA) model 
and mixtures of factor analyzers (MFA) 
play a valuable role in statistical data analysis, 
in particular, in cluster analysis, dimension reduction, 
and density estimation. 
Their usefulness was demonstrated in a wide range of applications 
in different fields such as bioinformatics \citep{J608}, 
informatics \citep{J607}, pattern recognition \citep{J606}, 
social and psychological sciences \citep{J609}, 
and environmental sciences \citep{J610}. 
The traditional formulation of the MFA model assumes that 
the latent component factors and errors 
jointly follow a multivariate normal distribution. 
However, in applied problems, the data will not always 
follow the normal distribution. 
To allow for clusters with heavy tails, 
\citet{J622} proposed the mixture of $t$-factor analyzers 
as a robust alternative to the MFA model, 
replacing the normality assumption of the factors and errors 
with a joint multivariate $t$-distribution.      
  
In recent times, a number of proposals have been developed 
to further generalize the MFA model to incorporate 
other non-normal distributions for the factors and/or errors. 
In particular, the past decade has seen mixtures models 
with skew component densities gaining increasing attention, 
being exploited as powerful tools for handling 
asymmetric distributional features in heterogeneous data. 
To name a few, there are 
mixtures of skew normal distributions \citep{J621,J004,J451}, 
mixtures of skew $t$-distributions \citep{J004,J027,J103,J164}, 
mixtures of the generalized hyperbolic family of distributions 
\citep{J375,J401}, and mixtures of other members of 
the skew elliptical class of distributions \citep{J066}. 
In almost all of these models, the skew component densities 
have the same or similar form as the skew normal distribution 
proposed by \citet{J001}. Their skewness is regulated by 
a vector of skewness parameters multiplied by 
a common skewing variable 
in its convolution-type characterization. 
An implication of this type of formulation is that 
skewness is assumed to be concentrated along a single direction 
in the feature space \citep{J360b} 
and hence referred to as a restricted skew distribution 
by \citet{J105}. 
Another formulation of skew distributions is developed 
by \citet{J002}, and referred to as 
the unrestricted skew distribution by \citet{J105}, 
since it does not rely on a common skewing variable. 
It thus allows for skewness to be in more than one direction,  
although each direction is parallel to 
the axes of the features space. 
Note that the restricted model 
is not nested within the unrestricted model. 
They are, however, identical in the univariate case. 
The unrestricted skew normal and skew $t$-distributions 
were adopted for the mixture models considered by 
\citet{J621,J027} and \citet{J103}. 
More recently, \citet{J164} considered the so-called 
canonical fundamental skew $t$ (CFUST) distribution 
as components in their mixture models, 
a more general skew distribution that encompasses 
both the restricted and unrestricted formulations 
of skew distributions. 
The CFUST distribution is a member of 
the class of canonical fundamental skew symmetric 
(CFUSS) distributions proposed by \citet{J008}. 
As the CFUSS distribution has a matrix of skewness parameters, 
it can flexibly handle multiple arbitrary directions of skewness. 
Another member of the class of CFUSS distributions is 
the canonical fundamental skew hyperbolic (CFUSH) distribution, 
which was studied recently by \citet{J616} and \citet{J631} 
using different names for this distribution.     
A detailed treatment of skew distributions 
can be found in \citet{B001}, \citet{J017},
\citet{J105}, and \citet{B025}.

A factor-analytic analogue of some of the above mentioned 
skew mixture models has been considered in other works, 
including a skew normal factor analysis model by \citet{J611}, 
a mixture of skew normal factor analyzers (MSNFA) by \citet{J159b}, 
a skew $t$-factor analysis by \citet{J160b}, 
its mixture model version (MSTFA) by \citet{J612}, 
a mixture of (generalized hyperbolic) skew $t$-factor analyzers 
(MGHSTFA) by \citet{J618}, and 
a mixture of generalized hyperbolic factor analzyers (MGHFA) 
by \citet{J620}. 
There are distinct differences between these models, 
not only in the choice of component densities, 
but also on where the assumption of skewness 
is placed in the model 
(that is, whether it is assumed 
for the factors and/or for the errors). 
These will be discussed later in this paper. 
However, a point of interest is that 
the vast majority of these models 
adopt the restricted form of skew distributions 
and hence share the same limitation that 
the component densities are designed for 
modelling skewness concentrated in a single direction. 
\citet{J623} have recently considered a MFA model 
with the component errors following 
the unrestricted skew $t$-distribution. 
Such a model is suitable for the case where skewness 
is exhibited along the directions of the feature axes. 
More recently, an MFA model based on the CFUSH distribution 
has been considered by \citet{J615}. 
Such a model embeds the so-called 
canonical fundamental skew normal (CFUSN) distribution 
and the unrestricted and restricted skew normal distributions 
as limiting cases.

In this paper, we propose a mixture of skew factor analyzers,  
adopting a CFUSS distribution 
as the joint distribution for the component factors and errors. 
For simplicity, we focus on the scale mixture of CFUSN (SMCFUSN) distribution. 
This new generalization of the MFA model can capture 
multiple directions of skewness  simultaneously 
while performing implicit dimension reduction.   
The proposed mixture of CFUSS factor analzyers (CFUSSFA) and 
the mixture of SMCFUSN factor analzyers (SMCFUSNFA) 
also formally encompass the mixtures of skew normal, 
skew $t$, and the mixture of CFUSSH factor analyzers 
by \citet{J159b}, \citet{J612}, 
and \citet{J623,J615}, respectively. 
For illustration, we shall focus on the $t$-distribution member 
of the CFUSS and SMCFUSN families of distributions, 
namely the canonical fundamental skew $t$ (CFUST) distribution, 
as it is one of the more commonly used distributions. 
However, it should be noted the same methodology 
can be applied to other members of the class of CFUSS distributions. 
For parameter estimation, an expectation--maximization (EM) algorithm 
\citep{J034} 
is implemented to compute the maximum likelihood (ML) estimates 
of the parameters in the model. 
Factor scores can be obtained as part of the EM algorithm.

The rest of this paper is organised as follows. 
Section \ref{s:back} provides a brief outline of the MFA model 
and the CFUSS distribution. We then examine and discuss 
the relationships between various existing skew factor models. 
In Section \ref{s:main}, we introduce the CFUSSFA model 
and present some of its nested cases. 
We then focus on the CFUSTFA model 
and implement an EM-type algorithm for parameter estimation in Section \ref{s:EM}. 
Implementation details are described in Section \ref{s:impl}.  
To demonstrate the usefulness of the proposed methodology, 
the CFUSTFA model is applied to 
two real datasets 
in Section \ref{s:real}.
Finally, concluding remarks are given in Section \ref{s:concl}.

\section{\large Background and related work}
\label{s:back}

\subsection{Notation}
\label{s:prelim}

We begin by establishing some notation to be used 
throughout this paper. 
Let $\bY$ denote a $p$-dimensional random vector. 
We also let $\bOne_p$ be a $p\times 1$ vector of ones, 
$\bI_p$ be the $p$-dimensional identity matrix, 
$\bJ_p$ be the $p\times p$ matrix of ones, 
and $\bzero$ be a vector/matrix of appropriate size. 
The operator $\mbox{diag}(\cdot)$, depending on the context, 
produces either a diagonal matrix with the specified elements 
or a vector containing the diagonal elements of a diagonal matrix. 
The notation $|\by|$ implies 
taking the absolute value of each element of $\by$.      

The probability density function (pdf) 
and cumulative distribution function (cdf) 
of the $p$-dimensional normal distribution 
with mean $\bmu$ and covariance $\bSigma$ 
are denoted by $\phi_p(\cdot; \bmu, \bSigma)$ 
and $\Phi_p(\cdot; \bmu, \bSigma)$, respectively, 
and the distribution itself is denoted by $N_p(\bmu, \bSigma)$. 
Analogously, the pdf and cdf of 
a $p$-dimensional $t$-distribution 
with $\nu$ degrees of freedom 
are denoted by $t_p(\cdot; \bmu, \bSigma, \nu)$ 
and $T_p(\cdot; \bmu, \bSigma, \nu)$, respectively. 
When $p=1$, the subscript $p$ 
will be dropped for convenience of notation. 
The notation $TN_p(\cdot; \mathbb{R}^+)$ 
and $Tt_p(\cdot; \mathbb{R}^+)$ denote 
the truncated normal and $t$-distributions, respectively, 
that are confined to the positive hyperplane.

\subsection{The class of CFUSS distributions}
\label{sec:CFUSS}

In many applications, the data or the clusters within the data
 are not symmetrically distributed. 
In this paper, we consider a flexible generalization 
of the MFA and M$t$FA models by adopting 
the CFUSS distribution \citep{J008} 
for the joint distribution of the factors and errors. 
We begin by examining 
the fundamental skew distribution,  
one of the more general formulations of skew distributions. 
Its density can be expressed as 
the product of a symmetric density and a skewing function. 
Formally, the density of $\bY$, 
a $p$-dimensional random vector 
following a CFUSS distribution, 
is given by
\begin{eqnarray}
	f(\by; \btheta) &=&	2^{r} f_p(\by; \btheta) \, Q_r(\by; \btheta),  
\label{CFUSS}
\end{eqnarray}  
where $f_p(\by; \btheta)$ is a symmetric density on $\mathbb{R}^p$, 
$Q_r(\by; \btheta)$ is a skewing function 
that maps $\by$ into the unit interval, 
and $\btheta$ is the vector 
containing the parameters of $\bY$. 
Let $\bU$ be a $r\times 1$ random vector, 
where $\bY$ and $\bU$ follow a joint distribution 
such that $\bY$ has marginal density $f_p(\by; \btheta)$ 
and $Q_r(\by; \btheta) = P(\bU > \bzero \mid \bY = \by)$. 
If the latent random vector $\bU$ 
has its canonical distribution
(that is, with mean $\bzero$ 
and scale matrix $\bI_r$), 
we obtain the canonical form of (\ref{CFUSS}), 
namely the CFUSS distribution.  
The class of CFUSS distributions encapsulates many existing distributions, 
including most of those mentioned earlier in this paper. 
We shall consider some particular cases 
of the class of CFUSS distributions here.

\subsubsection{The CFUSN distribution} 
\label{sec:CFUSN}

The skew normal member of the class of CFUSS distributions is 
the canonical fundamental skew normal (CFUSN) distribution. 
This can be obtained by taking $f_p$ to be a normal density, 
leading to $Q_r$ being a normal cdf. 
It follows that the density of the CFUSN distribution 
is given by
\begin{eqnarray}
f_{\mbox{\tiny{CFUSN}}}(\by; \bmu, \bSigma, \bDelta) 
	&=& 2^{r} \phi_q(\by; \bmu, \bOmega) 	\; 
	\Phi_r\left(\bDelta^T\bOmega^{-1}(\by-\bmu); 
	\bzero, \bLambda \right), 
\label{CFUSN}
\end{eqnarray}
where $\bOmega = \bSigma + \bDelta\bDelta^T$ and 
$\bLambda = \bI_r - \bDelta^T \bOmega^{-1} \bDelta$. 
In the above, 
$\bmu$ is a $p\times 1$ vector of location parameters,   
$\bSigma$ is a $p\times p$ positive definite scale matrix, and 
$\bDelta$ is a $p\times r$ matrix of skewness parameters.  
We shall adopt the notation 
$\bY \sim \mbox{CFUSN}_{p,r}(\bmu, \bSigma, \bDelta)$ 
if $\bY$ has the density given by (\ref{CFUSN}). 
Note that when $\bDelta = \bzero$, 
we obtain the (multivariate) normal distribution. 
In addition, a number of skew normal distributions 
are nested within the CFUSN distribution, 
including the version proposed by \citet{J001} 
and the version proposed by \citet{J002}. 
We shall follow the terminology of \citet{J105} 
and refer to them as the restricted and 
unrestricted skew normal distribution, respectively.  
\newline

\noindent
It is of interest to note that $\bY$ admits 
a convolution-type stochastic representation 
that facilitates the derivation of properties 
and parameter estimation via the EM algorithm. 
This is given by
\begin{eqnarray}
\bY &=& \bmu + \bDelta |\bU| + \be,
\label{CFUSN-conv}
\end{eqnarray} 
where $\bU$ follows a standard $r$-dimensional normal distribution, 
independently of $\be \sim N_p(\bzero, \bSigma)$. 
Hence, $|\bU|$ has a standard half-normal distribution.

\subsubsection{Scale mixture of CFUSN distributions} 
\label{sec:SMCFUSN}

In the next two subsections, we shall consider 
two skew distributions that were recently employed 
by \citet{J164} and \citet{J616} for their mixture models, 
namely the CFUST and HTH distributions, respectively. 
They are special cases of the class of the CFUSS distributions that can be 
obtained as a scale mixture of the CFUSN (SMCFUSN) distribution. 
By a normal scale mixture, we mean a distribution that 
can be defined by the stochastic representation 
\begin{eqnarray}
	\bY &=& \bmu + W^{\frac{1}{2}} \bY_0, 
\label{SM}
\end{eqnarray} 
where $\bY_0$ follows a central CFUSN distribution 
and $W$ is a positive (univariate) random variable 
independent of $\bY_0$. Thus, conditional on $W = w$, 
the density of $\bY$ is a CFUSN distribution 
with scale matrix ${w}\bSigma$. 
It follows that the marginal density of $\bY$ is given by
\begin{eqnarray}
f_{\mbox{\tiny{SMCFUSN}}} 
	(\by; \bmu, \bSigma, \bDelta; F_{\bzeta}) 
	&=&
	2^{r} \int_0^\infty 
	\phi_p \left(\by; \bmu, {w}\bOmega\right) \,
	\Phi_r\left(\frac{1}{\sqrt{w}}\bDelta^T\bOmega^{-1}(\by-\bmu); 
	\bzero, \bLambda \right) 
	dF_{\bzeta}(w), 
\nonumber\\ \label{SMSN}
\end{eqnarray}
where $F_{\bzeta}$ denotes the distribution function of $W$ 
indexed by the parameter $\bzeta$.  
We shall use the notation
$\bY \sim SMCFUSN_{p,r}(\bmu, \bSigma, \bDelta; F_{\bzeta})$ 
if the density of $\bY$ can be expressed in the form of (\ref{SMSN}).  
The class of SMCFUSN distributions is 
a generalization of the scale mixture of 
skew normal (SMSS) distributions considered by \citet{J066}.  
The latter adopts a restricted skew normal distribution 
in place of the CFUSN distribution here. 
This class can be obtained 
from the SMCFUSN distribution by taking $r=1$ 
(after reparameterization). 
Some special cases of the SMCFUSN distribution 
are listed in Table~\ref{tab:SMCFUSN}.

\begin{table*}
	\centering
	\hspace*{-1cm} 
		\begin{tabular}{|c||c|c|c|c|}
			\hline
			Model	&	Notation	&	Scaling density 
			& Symmetric density 	& Skewing function \\
			\hline \hline
			skew hyperbolic$^*$ &	CFUSH$^*$	&	$GIG(\psi,\chi,\lambda)$ & symmetric GH & symmetric GH \\
			skew $t$ &	CFUST	&	$\mbox{invGamma}(\frac{\nu}{2},\frac{\nu}{2})$ & $t$ & $t$ \\
			skew normal &	CFUSN	&	1 & normal & normal \\
			$t$ &	$t$	&	$\mbox{invGamma}(\frac{\nu}{2},\frac{\nu}{2})$ & $t$ & 1 \\
			normal &	N	&	1 & normal & 1 \\
			\hline
		\end{tabular}
	\hspace*{-1cm} 
	\caption{Some special cases of the scale mixture of CFUSN distributions.
	We let CFUS denote the canonical fundamental skew distribution, 
	SH the specialized hyperbolic distribution, 
	and invGamma the inverse Gamma distribution.  
	$^*$The CFUSH distribution is not identifiable  
	and hence \citet{J616} and \citet{J631} imposed different constraints on the parameters to achieve identifiability.  }
	\label{tab:SMCFUSN}
\end{table*}   

\subsubsection{The CFUSH distribution} 
\label{sec:CFUSH}

If the latent variable $W$ in (\ref{SM}) follows 
a generalized inverse Gaussian (GIG) distribution \citep{B104}, 
we obtain the canonical fundamental skew hyperbolic (CFUSH) distribution. 
In this case, the symmetric density $f_p$ in (\ref{CFUSS}) 
is a symmetric GH distribution $h_p(\cdot)$ and the skewing function 
becomes the cdf of a symmetric GH distribution $H_r(\cdot)$. 
The GIG density can be expressed as 
\begin{eqnarray}
f_{\mbox{\tiny{GIG}}} (w; \psi, \chi, \lambda) 
	&=&	\frac{\left(\frac{\psi}{\chi}\right)^{\frac{\lambda}{2}} w^{\lambda -1}}
		{2 K_\lambda(\sqrt{\chi\psi})} 
		e^{-\frac{\psi w + \frac{\chi}{w}}{2}},
\label{GIG}
\end{eqnarray}
where $W > 0$, the parameters $\psi$ and $\chi$ are positive, 
and $\lambda$ is a real  parameter. 
In the above, $K_\lambda(\cdot)$ denotes 
the modified Bessel function of the third kind of order $\lambda$. 
The density of a $p$-dimensional symmetric generalized hyperbolic distribution is given by
\begin{eqnarray}
	h_p(\by; \bmu, \bSigma,\bpsi,\chi,\lambda)
&=&	\left(\frac{\chi+\eta}{\psi}\right)^{\frac{\lambda}{2}-\frac{p}{4}}
\frac{\left(\frac{\psi}{\chi}\right)^{\frac{\lambda}{2}} 
	K_{\lambda-\frac{p}{2}}(\sqrt{(\chi+\eta)\psi})}
{(2\pi)^{\frac{p}{2}} |\bSigma|^{\frac{1}{2}} K_\lambda(\sqrt{\chi\psi})}.
\label{eq:SGH}
\end{eqnarray}  
\newline

\noindent
It is well known that the GH distribution has an identifiability issue 
in that the parameter vectors $\btheta=(\bmu, c\bSigma, c\psi, \chi/c, \lambda)$
and $\btheta^*=(\bmu, \bSigma, \psi, \chi, \lambda)$ both yield 
the same symmetric GH distribution (\ref{eq:SGH}) for any $c>0$. 
It is therefore not surprising that the CFUSH distribution also suffers from such an issue. 
To handle this, restrictions are imposed on some of the parameters of the CFUSH distribution. An example is the HTH distribution considered by \citet{J616}, 
where the constraint $\psi=\chi=\omega$ is used, 
leading to the density
\begin{eqnarray}
&& \hspace*{-0.8cm} f_{\mbox{\tiny{CFUSSH}}}
	(\by; \bmu, \bSigma, \bDelta, \omega, \lambda) 
	\nonumber\\ 
&& \hspace{-0.3cm} 
	= 2^r h_p\left(\by; \bmu, \bOmega, \omega, \omega, \lambda\right) 	
	H_r\left(\bDelta^T\bOmega^{-1}(\by-\bmu) 	
	\left(\frac{\omega}{\omega+\eta}\right)^{\frac{1}{4}}; 
	\bzero, \bLambda, \lambda-\textstyle\frac{p}{2}, 
	\gamma, \gamma\right), \nonumber\\
\label{HTH}
\end{eqnarray} 
where $\gamma = \sqrt{\psi(\omega+\eta)}$. 
This particular parameterization is refer to as
the canonical fundamental skew specialized hyperbolic (CFUSSH) distribution.
Note that in their terminology, they are using `hidden truncation' 
to describe the latent skewing variable that 
follows a truncated distribution 
in the convolution-type characterization of the CFUSH distribution. 
Another alternative is to restrict the parameters of $W$ so that, for example, $E(W)=1$.  
A commonly used constraint on the GH distribution is to set $|\bSigma|=1$. 
This can be applied to the CFUSH distribution to achieve identifiability; 
see also the unrestricted skew normal generalized hyperbolic (SUNGH) distribution considered by \citet{J631}.

\subsubsection{The CFUST distribution} 
\label{sec:CFUST}

The CFUST distribution is the skew $t$-distribution 
member of the class of CFUSS distributions, where the symmetric distribution 
is taken to be a (multivariate) $t$-distribution. 
This can be obtained by letting $\frac{1}{W}$ 
be a random variable that has a 
$\mbox{gamma}(\frac{\nu}{2}, \frac{\nu}{2})$ distribution. 
Thus, its density is given by 
\begin{eqnarray}
&& f_{\mbox{\tiny{CFUST}}}(\by; \bmu, \bSigma, \bDelta, \nu) \nonumber\\
	=&& 2^r t_p(\by; \bmu, \bOmega, \nu) 	
	T_r\left(\bDelta^T\bOmega^{-1}(\by-\bmu); 
	\bzero, \left(\frac{\nu+\eta}{\nu+p}\right) \bLambda, 
	\nu+p\right). \nonumber\\
\label{CFUST}
\end{eqnarray}
We shall adopt the notation 
$\bY \sim \mbox{CFUST}_{p,r}(\bmu, \bSigma, \bDelta, \nu)$ 
if $\bY$ has the density given by (\ref{CFUST}). 
\newline

\noindent
The CFUST distribution can be represented by 
a number of stochastic representations, 
including the convolution of a half $t$-random vector $|\bU|$ 
and a $t$-random vector $\be$, given by
\begin{eqnarray}
\bY &=& \bmu + \bDelta |\bU| + \be,
\label{convolution}
\end{eqnarray} 
where $\bU$ and $\be$ have a joint $t$-distribution  given by
\begin{eqnarray}
\left[\begin{array}{c} \bU \\ \be \end{array}\right] 
	&\sim& t_{r+p} 
	\left(\left[\begin{array}{c} \bzero \\ \bzero \end{array}\right], 
	\left[\begin{array}{cc} \bI_r & \bzero\\ \bzero & \bSigma 
	\end{array}\right], \nu \right). \nonumber
\label{convolution2}
\end{eqnarray}
From (\ref{convolution}), we can obtain 
the mean and covariance matrix $\bX$, which are given by 
\begin{eqnarray} 
E(\bY) &=& \bmu + a(\nu) \bDelta \bOne_r \nonumber
\end{eqnarray}
and
\begin{eqnarray}
\mbox{cov}(\bY) &=& (\frac{\nu}{\nu-2}) 
	\left[\bSigma + \left(1-\frac{2}{\pi}\right)\bDelta\bDelta^T\right]
	+ \left[\frac{2\nu}{\pi (\nu-2)} + a(\nu)^2\right] \bDelta \bJ_r \bDelta^T, 
	\nonumber
\end{eqnarray}
where $a(\nu) = \sqrt{\frac{\nu}{2}} \Gamma(\frac{\nu-1}{2}) 
\left[\Gamma(\frac{\nu}{2})\right]^{-1}$. 
\newline

In addition to the CFUSN distribution (and its nested special/limiting cases), 
the CFUST distribution embeds a number of commonly used distributions 
as special or limiting cases. 
This includes the unrestricted $t$-distribution by \citet{J002} 
(obtained by taking $\bDelta$ to be a diagonal $p\times p$ matrix, 
and letting $\nu \rightarrow \infty$ for the skew normal case), 
the restricted skew $t$-distributions (obtained by setting $r=1$), 
and the $t$-distribution (obtained by setting $\bDelta=0$). 
Concerning the identifiability of the CFUST model, 
it can be observed from (\ref{convolution}) that 
it bears a resemblance to the FA model (\ref{FA}). 
Indeed, it can be viewed as a FA model with latent factors 
following a half $t$-distribution and the skewness matrix 
acting as the factor loading matrix. 
However, unlike the FA model, the term $\bDelta|\bU|$ 
in the CFUST distribution is not rotational invariant. 
However, it is invariant to permutations of the columns of $\bDelta$, 
but this does not affect the number of free parameters in the CFUST model.

\subsection{Factor analysis (FA) and mixture of factor analyzers (MFA)}

The factor analysis (FA) model postulates that 
the correlations between the variables in $\bY$ 
can be explained by the linear dependence of $\bY$ 
on a lower-dimensional latent factor $\bX$, as given by 
\begin{eqnarray}
\bY = \bmu + \bB \bX + \bepsilon,
\label{FA}
\end{eqnarray}
where $\bB$ is a $p\times q$ matrix of factor loadings, 
$\bX$ is a $q$-dimensional latent factor ($q\leq p$), 
and $\bepsilon$ is a $p\times 1$ vector of error variables. 
In the traditional case of a normal MFA model, 
it is assumed that $\bX \sim N_q(\bzero, \bI_q)$ 
and $\bepsilon \sim N_p(\bzero, \bD)$, 
and that they are independently distributed of each other. 
Also, $\bD$ is taken to be a diagonal matrix 
with diagonal elements given by $\bd$; that is, 
$\bD = \mbox{diag(\bd)}$. 
This follows from the assumption that the variables 
in $\bY$ are distributed independently after allowing for the factors. 
From (\ref{FA}), the marginal density of $\bY$ 
is given by $N_p(\bmu,\bB\bB^T+\bD)$. 
In the case of multivariate latent factors (that is, $q>1$), 
the FA model suffers from an identifiability issue 
due to the rotational invariance of $\bB\bU$. 
To ensure the FA model can be uniquely defined, 
$q(q-1)/2$ constraints can be imposed 
on the factor loadings $\bB$ 
to reduce the number of free parameters 
for the covariance matrix 
from $\frac{1}{2}p(p+1)$ to $p + pq - \frac{1}{2}q(q-1)$. 
\newline

\noindent
The mixture of factor analzyers (MFA) model \citep{J617,B003} 
is a mixture version of the FA model wherein, 
given that $\bY$ belongs to the $i$th component of the mixture model, 
it can be the expressed in the form of (\ref{FA}). 
The density of MFA is given by
\begin{equation}
f(\by; \bPsi) = \sum_{i=1}^g \pi_i f_i(\by; \btheta_i),
\label{mix}
\end{equation}   
where the $\pi_i$ $(i=1,\ldots,g)$ denote the mixing proportions, 
which are non-negative and sum to one. 
The generic function $f_i(\cdot)$ denotes the density 
of the $i$th component of the mixture model 
with parameters $\btheta_i$. 
In the case of the classical (normal) MFA model, 
this comprises the mean vectors $\bmu_i$, 
the loading matrices $\bB_i$, 
and the scale matrices $\bD_i$. \newline

\noindent
\citet{J622} proposed the mixture of 
$t$-factor analysers (M$t$FA) 
as a more robust version of the MFA model. 
It is defined in a similar way to (\ref{FA}), 
but assuming that the factors and the errors 
jointly follow a multivariate $t$-distribution. 
More formally, we have   
\begin{eqnarray}
\bY_j = \bmu_i +  \bB_i\bX_{ij} + \bepsilon_{ij}, 
    \hspace{1cm} \mbox{with probability }\pi_i, 
\label{MtFA}
\end{eqnarray}
where 
\begin{eqnarray}
\left[\begin{array}{c} \bX_{ij} \\ \bepsilon_{ij} \end{array}\right] 
	&\sim& t_{q+p} \left(
	\left[\begin{array}{c} \bzero \\ \bzero \end{array}\right], 
	\left[\begin{array}{cc} \bI_q & \bzero\\ \bzero & \bD_i 
	\end{array}\right], \nu_i \right).
\label{tFA}
\end{eqnarray}
In this case, the marginal density of $\bY_j$ is 
$\sum_{i=1}^g \pi_i \, t_p(\bmu_i,\bB_i\bB_i^T+\bD_i, \nu_i)$. 
The M$t$FA model has the same identifiability problem 
as the MFA model and thus the same constraints 
on the factor loadings can be imposed.     
\newline

\subsection{Related models}
\label{s:skewFA}

As discussed in the aforementioned sections, 
there are a number of existing proposals 
for skew factor analysis models 
or mixtures of skew factor analyzers. 
They differ in 
(i) how the adopted skew distribution is characterized 
and (ii) whether it is the factors or the errors (or both) 
that are assumed to follow the chosen skew distribution. 
In this discussion, we shall focus on the more relevant models 
that are mentioned in Section \ref{intro}. 

Concerning (ii), for the case of a single factor analysis model, 
\citet{J611} considered the restricted skew normal distribution 
for its factors. \citet{J624} proposed the so-called 
generalised skew normal factor model 
which is equivalent to a FA model 
with the errors following a CFUSN distribution.    
In the case of mixtures of skew factor analyzers, 
we note that for the MSNFA and MSTFA models \citep{J159b,J612} 
the factors follow a (restricted) skew normal 
and skew $t$-distribution, respectively. 
In contrast, for the MGHSTFA and MGHFA models \citep{J618,J620}, 
the errors are assumed to follow a GHST distribution 
and a (special case) of 
the generalized hyperbolic (GH) distribution, respectively. 
Similarly, the model by \citet{J623} assumes that 
the errors follow the skew $t$-distribution by \citet{J002}, 
which we shall refer to 
as the unrestricted skew $t$-factor analyzers (MuSTFA) model 
(following the terminology of \citet{J105}) 
to distinguish it from the other skew $t$-factor analyzers models 
considered in this paper. The more recent HTHFA model 
proposed by \citet{J615} assumes that the factors 
follow a CFUSSH distribution and the errors marginally 
follow a hyperbolic distribution.
As discussed previously, we shall henceforth refer to it as the CFUSSHFA model.    

Concerning (i), 
it should be noted that although also called 
the skew $t$-distribution by \citet{J618},
the GHST distribution is different from
the restricted skew $t$-distribution. 
The former arises as a special case of the GH distribution 
and so exhibits different tail behaviour 
to the restricted skew $t$-distribution. 
Moreover, it does not incorporate a skew normal distribution 
as it becomes the (symmetric) normal distribution 
as the degrees of freedom approach infinity. 
The GH distribution adopted by the GHFA model 
has restrictions placed on some of its parameters 
(similar to that for the CFUSSH distribution)
due to an identifiability issue. 
For ease of reference, 
a summary of the above mentioned MFA models 
is listed in Table \ref{tab:sum}. 
Note that for brevity this list is not exhaustive 
and only the most relevant models are included.

A detailed study of the analytical differences 
between these distributions is beyond the scope of this paper. 
However, it is of interest here to recognize that 
although formulated differently, 
the skewness in these approaches 
(with the exception of uMSTFA and CFUSSHFA) 
is regulated by a single latent skewing variable and thus, 
in effect, is somewhat similar to the special case of $r=1$ 
of the proposed CFUSSFA and CFUSTFA models.  
This implies realizations of the skewing variable 
are confined to lie about a line in the feature space 
and therefore are limited to modelling skewness 
concentrated along a single direction \citep{J360b}. 
In the case of the uMSTFA model, 
there are $q$ skewing variables that are uncorrelated 
and taken to be feature-specific.  
On the other hand, the CFUSS distribution 
allows for $r$ latent skewness variables 
which enables it to represent skewness 
along multiple arbitrary directions.  

\begin{figure}
	\centering
		\includegraphics[width=0.85\textwidth]{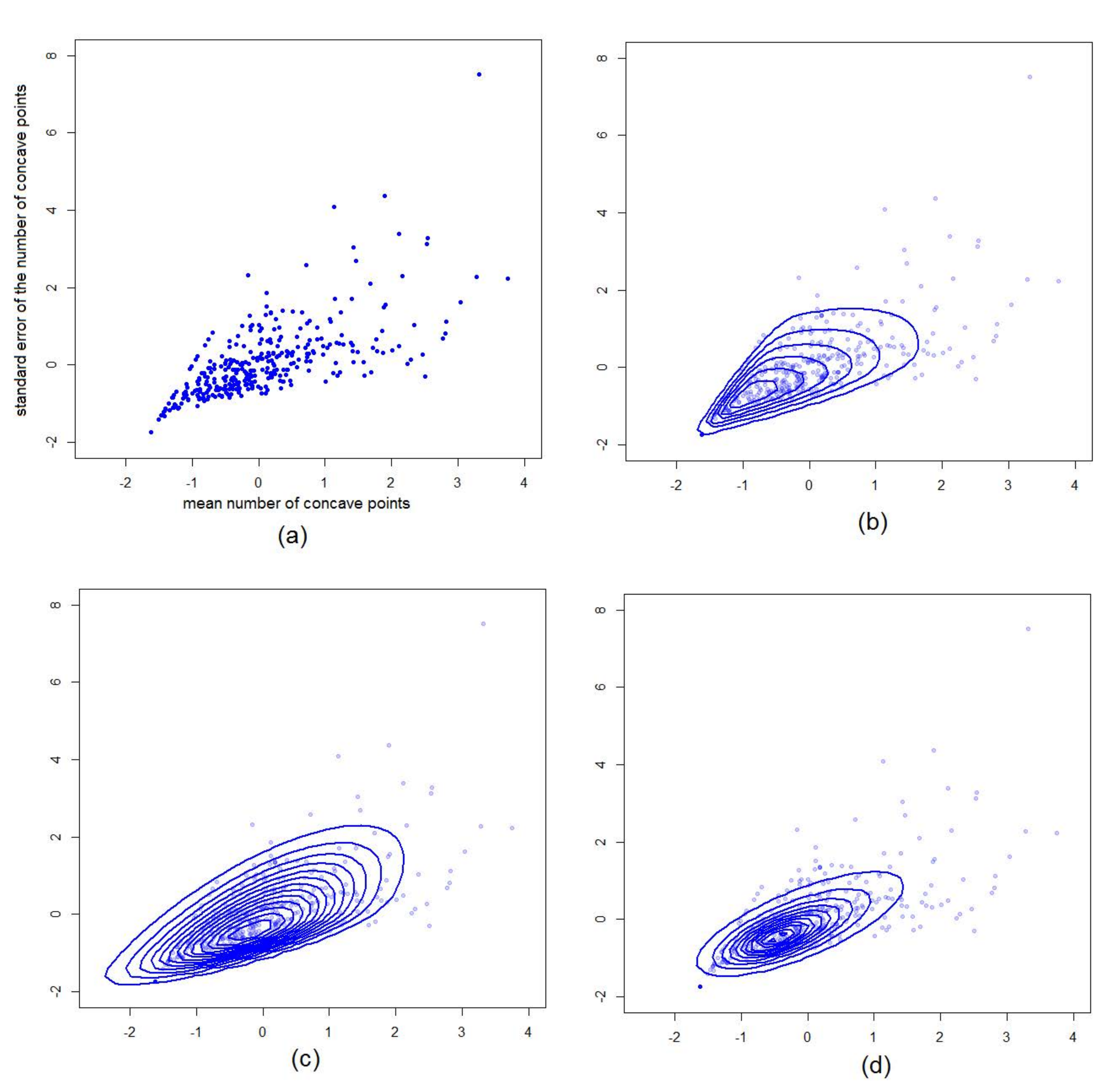}
	\caption{Modelling the WDBC data with three skew distributions. 
	(a) A scatterplot of benign observations on the two variables: 
	mean and standard error of the number of concave points. 
	(b) The contours of the density of the fitted CFUSN distribution.
	(c) The contours of the density of the fitted restricted SN distribution.
	(d) The contours of the density of the fitted generalized hyperbolic distribution.}
	\label{fig:shapes}
\end{figure}

The practical implications of this issue 
can be illustrated on 
the Wisconsin Diagnostic Breast Cancer (WDBC) dataset \citep{UCI}. 
Consider the subset consisting of two variables, 
namely, the mean number of concave points 
and the standard error of the number of concave points.   
A scatterplot of the observations from benign patients 
is shown in Figure \ref{fig:shapes}(a). 
The distribution of observations is 
apparently highly asymmetric and seems to exhibit skewness 
in two distinct directions. 
Upon fitting a restricted SN distribution to the data, 
we observed from Figure \ref{fig:shapes}(c) 
that it successfully captures one of the skewness directions 
but is having difficulty with the other 
(see the lower left corner of the Figure). 
The GH distribution also finds this situation 
challenging to model (Figure \ref{fig:shapes}(d)). 
The CFUSN distribution (Figure \ref{fig:shapes}(b)) 
provides a much closer fit 
to the data and is capable of modelling 
both directions of skewness (Figure \ref{fig:shapes}(a)).

\begin{table}
	\centering
	\hspace*{-1cm} 
		\begin{tabular}{|c||c|c|c|c|}
			\hline
			\textbf{Skew MFA Model} & \textbf{Notation} & \textbf{Factors} 
			& \textbf{Errors} & \textbf{References} \\
			\hline \hline
			Restricted skew normal 	&	MSNFA &	rMSN & normal & \citet{J159b} \\\hline 
			Restricted skew $t$ 		&	MSTFA &	rMST & $t$ & \citet{J612} \\\hline 
			Unrestricted skew $t$ 	&	uMSTFA & $t$ 		& uMST & \citet{J623} \\\hline 
			Generalized hyperbolic* 	&	MGHFA &	SGH 	& GH & \citet{J620} \\\hline  
			GH skew $t$ 						&	MGHSTFA &	$t$ & GHST & \citet{J618} \\\hline  
			Common GH skew $t$ 			&	MCGHSTFA &	GHST & $t$ & \citet{J619} \\\hline  
			CFUS hyperbolic*					&	CFUSHFA 	&	CFUSH & SH & \citet{J615} \\ 
								&	 	&	 &  & \citet{J631} \\\hline 
			CFUS symmetric					&	CFUSSFA 	&	CFUSS & symmetric & this paper \\\hline
			SMCFUSN									&	SMCFUSNFA 	&	SMCFUSN & SMN & this paper \\\hline
			CFUS normal							&	CFUSNFA 	&	CFUSN & normal & this paper \\\hline 
			CFUS hyperbolic				  &	CFUSHFA 	&	CFUSH & hyperbolic & this paper \\\hline    
			CFUS $t$ 									&	CFUSTFA 	&	CFUST & $t$ & this paper \\\hline 
		\end{tabular}
	\caption{Summary of different skew generalizations of 
	the mixtures of factor analyzers (MFA) models. 
	Here, we use the notation rMSN, rMST, uMST, SGH, GHST, SH, 
	CFUSS, CFUSN, SMCFUSN, SMN, CFUSH, and CFUST to refer to the (restricted) skew normal, 
	(restricted) skew $t$, unrestricted skew $t$, 
	symmetric generalized hyperbolic, generalized hyperbolic skew $t$, 
	specialized hyperbolic, canonical fundamental skew symmetric, 
	canonical fundamental skew normal, scale mixture of CFUSN, 
	scale mixture of normal, canonical fundamental skew hyperbolic, 
	and canonical fundamental skew $t$-distribution, respectively. 
	For brevity, we include only the most relevant models in this table. 
	*This distribution suffers from an identifiability issue and hence constraints were imposed on the parameters; see the cited references for examples of these constraints. }
	\label{tab:sum}
\end{table}

\section{\large Mixture of CFUSS factor analyzers (CFUSSFA) model}
\label{s:main}

Here, we propose to generalize the M$t$FA (\ref{MtFA}) model 
to the case where the factors and errors 
are jointly distributed as a CFUSS distribution. 
We replace the $t$-distribution in (\ref{tFA}) 
with a CFUSS distribution. 
For simplicity, we focus on the class of SMCFUSN distributions. 
Let $\bY_1, \ldots, \bY_n$ be a random sample of $n$ observations 
of $\bY$. Accordingly, the mixture of SMCFUSN factor analzyers 
(SMCFUSNFA) can be formulated as
\begin{eqnarray}
	\bY_j &=& \bmu_i + \bB_i \bX_{ij} + \bepsilon_{ij}, 
\label{SMCFUSNFA}
\end{eqnarray}
with probability $\pi_i \; (i=1, \ldots, g)$, 
where\begin{eqnarray}
\left[\begin{array}{c} \bX_{ij} \\ \bepsilon_{ij} \end{array}\right] 
	&\sim& \mbox{\textit{SMCFUSN}}_{q+p} 
	\left(\left[\begin{array}{c} \bzero \\ \bzero \end{array}\right], 
	 \left[\begin{array}{cc} \bI_q & \bzero\\ \bzero & \bD_i \end{array}\right], 
	\left[\begin{array}{c} \bDelta_i \\ \bzero \end{array}\right];  
	F_{\bzeta_i}\right). \nonumber\\
\label{SMCFUSNFA2}
\end{eqnarray} 
Some special cases of SMCFUSNFA shall be considered next.

\subsection{The CFUSN factor analysis (CFUSNFA) model}
\label{s:CFUSNFA}

The mixture of CFUSN factor analzyers (CFUSNFA) 
is a degenerate case of CFUSSFA. 
It can be formulated as 
\begin{eqnarray}
	\bY_j &=& \bmu_i + \bB_i \bX_{ij} + \bepsilon_{ij}, 
\label{CFUSNFA}
\end{eqnarray}
with probability $\pi_i \; (i=1, \ldots, g)$, 
where
\begin{eqnarray}
\left[\begin{array}{c} \bX_{ij} \\ \bepsilon_{ij} \end{array}\right] 
	&\sim& CFUSN_{q+p} 
	\left(\left[\begin{array}{c} \bzero \\ \bzero \end{array}\right], 
	 \left[\begin{array}{cc} \bI_q & \bzero\\ \bzero & \bD_i 
	\end{array}\right], 
	\left[\begin{array}{c} \bDelta_i \\ \bzero \end{array}\right]\right).
\label{CFUSNFA2}
\end{eqnarray}
Thus, marginally, the factors $\bX_{ij}$ $(j=1, \ldots, n)$ 
follow a standard $q$-dimensional CFUSN distribution, 
whereas the errors $\bepsilon_{ij}$ follow 
a $p$-dimensional normal distribution. 
When $\bDelta = \bzero$, we retrieve the MFA model. 
When $r=1$, we retrieve a skew normal MFA model 
equivalent to the MSNFA model proposed by \citet{J159b}. 
Note that in the formulation of the MSNFA model, 
the authors adopt a slightly different parametrization 
so that the factors have expected value being $\bzero$ 
and the covariance matrix is equal to the identity matrix.      

\subsection{The CFUSH factor analysis (CFUSHFA) model}
\label{s:CFUSHFA}

To obtain the mixture of CFUSH factor analzyers (CFUSHFA), 
let $F_{\bzeta_i}$ for the $i$th component 
denote the GIG distribution function
with density defined in (\ref{GIG}). 
Then $\bzeta_i$ contains the parameters 
$\psi_i$, $\chi_i$, and $\lambda_i$ for $i=1, \ldots, g$. 
The resulting model is a CFUSHFA model which is given by 
\begin{eqnarray}
	\bY_j &=& \bmu_i + \bB_i \bX_{ij} + \bepsilon_{ij}, 
\label{CFUSHFA}
\end{eqnarray}
with probability $\pi_i \; (i=1, \ldots, g)$, 
where
\begin{eqnarray}
\left[\begin{array}{c} \bX_{ij} \\ \bepsilon_{ij} \end{array}\right] 
	&\sim& \mbox{\textit{CFUSH}}_{q+p} 
	\left(\left[\begin{array}{c} \bzero \\ \bzero \end{array}\right], 
	 \left[\begin{array}{cc} \bI_q & \bzero\\ \bzero & \bD_i \end{array}\right], 
	\left[\begin{array}{c} \bDelta_i \\ \bzero \end{array}\right], 
	\psi_i, \chi_i, \lambda_i\right). \nonumber\\
\label{CFUSHFA2}
\end{eqnarray}
In this case, the marginal distribution of the factors $\bX_{ij}$ 
is a standard $q$-dimensional CFUSH distribution, 
whereas the errors $\bepsilon_{ij}$ follow 
a $p$-dimensional hyperbolic distribution.

\subsection{The CFUST factor analysis (CFUSTFA) model}
\label{s:CFUSTFA}

\noindent 
Consider now adopting the CFUST distribution 
for the joint distribution of the factors and errors 
in (\ref{SMCFUSNFA2}). This corresponds to 
the special case of the SMCFUSNFA model 
with $F_{\bzeta_i}$ being the inverse gamma distribution function
with parameter $\bzeta_i = \nu_i$. 
Henceforth, we shall refer to this model 
as the CFUST factor analysis (CFUSTFA) model. 
This model can be formulated as 
\begin{eqnarray}
	\bY_j &=& \bmu_i + \bB_i \bX_{ij} + \bepsilon_{ij}, 
\label{CFA}
\end{eqnarray}
with probability $\pi_i \; (i=1, \ldots, g)$, 
where
\begin{eqnarray}
\left[\begin{array}{c} \bX_{ij} \\ \bepsilon_{ij} \end{array}\right] 
	&\sim& CFUST_{q+p} 
	\left(\left[\begin{array}{c} \bzero \\ \bzero \end{array}\right], 
	 \left[\begin{array}{cc} \bI_q & \bzero\\ \bzero & \bD_i 
	\end{array}\right], 
	\left[\begin{array}{c} \bDelta_i \\ \bzero \end{array}\right], 
	\nu_i\right).
\label{CFUSTFA}
\end{eqnarray}
It is clear that, marginally, 
the factors $\bX_{ij}$ $(j=1, \ldots, n)$ 
follow a standard $q$-dimensional CFUST distribution, 
whereas the errors $\bepsilon_{ij}$ follow 
a $p$-dimensional $t$-distribution. 
More specifically, $\bX_{ij} \sim CFUST_q(\bzero, \bI_q, \bDelta_i, \nu_i)$ 
and $\bepsilon_{ij} \sim t_p(\bzero, \bD_i, \nu_i)$. 
Note that similar to the $t$FA model, 
the $\bX_{ij}$ and $\bepsilon_{ij}$ here
 are not independent but are uncorrelated. 
It follows from (\ref{CFA}) that 
the mean and covariance matrix of the factors are given by
\begin{eqnarray}
&&	E\left(\bX_{ij}\right) = a(\nu_i) \bDelta_i \bOne_r \nonumber\\
\mbox{and} 	&& \nonumber\\
\mbox{cov}\left(\bX_{ij}\right) &=& \left(\frac{\nu_i}{\nu_i-2}\right) 
		\left[\bI_q + \left(1-\frac{2}{\pi}\right) \bDelta_i\bDelta_i^T\right]
		+ \left[\frac{2}{\pi}\left(\frac{2}{\nu_i-2}\right) - a(\nu_i)^2\right] 
		\bDelta_i \bJ_r \bDelta_i^T,
\nonumber
\end{eqnarray}
respectively.  
\newline

\noindent
It follows from (\ref{CFA}) that 
the marginal density of $\bY_j$ is a CFUST distribution; 
that is, given that $\bY_j$ belongs to 
the $i$th component of the mixture model, it is distributed as
\begin{eqnarray}
\bY_j &\sim& \mbox{CFUST}_{p,r}(\bmu_i, \bB_i\bB_i^T+\bD_i, \bB_i\bDelta_i, \nu_i).
\label{Yden} 
\end{eqnarray}
Hence, the mean and covariance matrix of $\bY_j$ are given by
\begin{eqnarray}
E(\bY_j) &=& \bmu + a(\nu_i) \bB_i \bDelta_i \bOne_r, \nonumber
\end{eqnarray}
and
\begin{eqnarray}
 \mbox{cov}(\bY_j)  
 &=& \left(\frac{\nu_i}{\nu_i-2}\right) 
		\left[\bB_i\bB_i^T + \bD_i + \left(1-\frac{2}{\pi}\right)
		\bB_i\bDelta_i\bDelta_i^T\bB_i^T\right]
		\nonumber\\&& 
		+ \left[\frac{2}{\pi}\left(\frac{2}{\nu_i-2}\right) - a(\nu_i)^2\right]
		\bB_i \bDelta_i \bJ_r \bDelta_i^T \bB_i,
\end{eqnarray}
respectively. 
\newline

\noindent
Accordingly, the density of $\bY_j$ is a $g$-component CFUST mixture density, given by
\begin{eqnarray}
f(\by_j; \bPsi) 
&=& \sum_{i=1}^g \pi_i \, 
	f_{\mbox{\tiny{CFUST}}}
	\left(\by_j; \bmu_i, \bSigma_i^*, \bDelta_i^*, \nu_i\right), 
\label{FM}
\end{eqnarray}
where the vector $\bPsi = \left\{\pi_1, \ldots, \pi_{g-1}, 
\btheta_1^T, \ldots, \btheta_g^T\right\}^T$ 
contains all the unknown parameters of the mixture model 
and $\btheta_i$ is the vector of unknown parameters 
of the $i$th component of the mixture model, 
comprising the elements of $\bmu_i$, $\bDelta_i$, 
$\bB_i$, $\bD_i$, and $\nu_i$. 
In the above, we let $\bSigma^* = \bB_i\bB_i^T+\bD$ 
and $\bDelta^* = \bB_i \bDelta$ for notational convenience. 
It should be noted that $r$ need not be smaller than $q$.
However, for simplicity, we will focus our attention 
on the cases where $r\leq q$ in the applications in Section \ref{s:real}.  
In addition, note that the CFUSTFA model 
reduces to the $t$FA model when $\bDelta_i=\bzero$, 
and reduces to the FA model when $\bDelta_i=\bzero$
 and $\nu_i \rightarrow \infty$.
Concerning identifiability, 
note that in the case of a CFUSTFA model 
$\bX_{ij}$ is no longer rotational invariant 
due to its 
following a non-symmetrical distribution. 
Hence, the CFUSTFA model does not inherit the aforementioned issue 
concerning the $\bB_{i}$ in the FA and $t$FA models. 
\newline

\section{Parameter estimation for the SMCFUSNFA model}
\label{s:EM}

Parameter estimation can be carried out 
using the maximum likelihood (ML) approach 
via the EM algorithm. For the CFUSSFA model, 
we exploit a variant of the EM algorithm 
that is useful when the M-step 
is relativity difficult to compute. 
The expectation--conditional maximization (ECM) algorithm \citep{J630}
replaces the M-step with a sequence of 
computationally simpler conditional--maximization (CM) steps 
by conditioning on the preceding parameters being estimated. 
\newline

\noindent 
From (\ref{SMCFUSNFA}) and (\ref{SMCFUSNFA2}) 
and using (\ref{CFUSN-conv}), 
the SMCFUSNFA model admits a five-level hierarchical representation.  
By expressing $\bX_{ij}$ in terms of 
the latent variables $\bU_{ij}$ and $W_{ij}$, 
it follows that 
\begin{eqnarray}
\bY_j	\mid \bx_{ij}, w_{ij}, Z_{ij} =1	&\sim& 
	N_p\left(\bB_i\bx_{ij} + \bmu_i, {w_{ij}} \bD_i\right),
	\nonumber\\
\bX_{ij}	\mid \bu_{ij}, w_{ij}, Z_{ij} =1	&\sim& 
	N_q\left(\bDelta_i|\bu_{ij}|, {w_{ij}} \bI_q\right),
	\nonumber\\
|\bU_{ij}| \mid w_{ij}, Z_{ij} =1	&\sim&	
	TN_r\left(\bzero, {w_{ij}} \bI_r; \mathbb{R}^+\right), 
	\nonumber\\
W_{ij} \mid Z_{ij} =1	&\sim&	F_{\bzeta_i},
		\nonumber\\
\bZ_j &\sim& \mbox{Multi}_g(1; \bpi),
\label{hierarchical}
\end{eqnarray}
where 
$\mbox{Multi}_g(1; \bpi)$ denotes 
the multinomial distribution having $g$ categories 
with associated probabilities 
$\bpi = (\pi_1, \ldots, \pi_g)^T$. 
In the above, 
we let  
$\bZ_j = (Z_{1j},\, \ldots,\, Z_{gj})^T$ $(j=1,\ldots, n)$ 
be the vector of latent indicator variables, 
where $Z_{ij} = 1$ if $\by_j$ belongs to 
the $i$th component of the mixture model and $Z_{ij}=0$ otherwise. 
One may observe from (\ref{hierarchical}) that 
the last four levels are identical to that 
for a finite mixture of SMCFUSN distributions.     
\newline

Under the EM framework, the indicator labels $Z_{ij}$ 
and the latent variables $W_{ij}$, $\bU_{ij}$, 
and $\bX_{ij}$ are treated as missing data. 
Thus, the complete-data vector is given by 
$(\by^T, \bx^T, \bu^T, \bw^T, \bz^T)^T$,
where $\by = (\by_1^T, \ldots, \by_n^T)^T$, 
$\bx = (\bx_{11}^T, \ldots, \bx_{gn}^T)^T$, 
$\bu = (\bu_{11}^T, \ldots, \by_{gn}^T)^T$,  
$\bw = (w_{11}^T, \ldots, w_{gn}^T)^T$, and 
$\bz = (\bZ_1^T, \ldots, \bZ_n^T)^T$. 
The log likelihood function and the $Q$-function 
can be derived using (\ref{hierarchical}). 
Accordingly, the complete-data log likelihood function is given by  
\begin{eqnarray}
\log L_c(\bPsi) &=& \sum_{i=1}^g \sum_{j=1}^n Z_{ij} 
		\left[\log \pi_{ij} 
		-\frac{1}{2w_{ij}} 
		\left(\bx_{ij}-\bDelta_i|\bu_{ij}|\right)^T  
		\left(\bx_{ij}-\bDelta_i|\bu_{ij}|\right)
		+ \log f_{\bzeta_i}(w_{ij})
	\right. \nonumber\\&& 
		-\frac{1}{2w_{ij}} 
		\left(\by_j -\bmu_i - \bB_i\bx_{ij}\right)^T \bD_i^{-1}
		\left(\by_j -\bmu_i - \bB_i\bx_{ij}\right)	
		-\frac{1}{2}\log|\bD_i|
			\Big] 
\label{logL}
\end{eqnarray}
where additive constants and terms that 
do not involve any parameters of the model are omitted.   

\subsection{E-step}
\label{s:E}

On the E-step, we compute the so-called $Q$-function, 
which is the conditional expectation of 
the complete-data log likelihood function 
given the observed data using 
the current estimates of the model parameters. 
Let the superscript $(k)$ on the parameters 
denote the updated estimates after the $k$th iteration 
of the EM algorithm. To compute the $Q$-function, 
we need to evaluate the following conditional expectations:
\begin{eqnarray}
z_{ij}^{(k)} &=&	
	E_{\Psi^{(k)}} \left[Z_{ij}=1 \mid \by_j\right] \label{E0},\\
w_{ij}^{(k)}	&=&	
	E_{\Psi^{(k)}} \left[\frac{1}{W_{ij}} \mid \by_j, Z_{ij}=1\right], 		\label{E1}\\
\be_{1ij}^{(k)}	&=&	E_{\Psi^{(k)}} \left[\frac{1}{W_{ij}} |\bU_{ij}|
		\mid \by_j, Z_{ij}=1 \right], 	\label{E2}\\
\be_{2ij}^{(k)} &=&	E_{\Psi^{(k)}} \left[\frac{1}{W_{ij}} |\bU_{ij}| |\bU_{ij}|^T
		\mid \by_j, Z_{ij}=1 \right],	\label{E3}\\	
\be_{3ij}^{(k)}	&=&	E_{\Psi^{(k)}} \left[\frac{1}{W_{ij}} \bX_{ij}
		\mid \by_j, Z_{ij}=1 \right], 	\label{E4}\\
\be_{4ij}^{{(k)}} &=&	
		E_{\Psi^{(k)}} \left[\frac{1}{W_{ij}} \bX_{ij} \bX_{ij}^T
		\mid \by_j, Z_{ij}=1 \right],	\label{E5}\\	
\be_{5ij}^{(k)} &=&	
		E_{\Psi^{(k)}} \left[\frac{1}{W_{ij}} \bX_{ij} |\bU_{ij}|^T
		\mid \by_j, Z_{ij}=1 \right].	\label{E6}		
\end{eqnarray}

\noindent
The exact expressions for these conditional expectations 
will depend on the form of $F_{\bzeta}$. 
In addition, it should be noted that 
extra conditional expectations may be needed 
for the CM-steps related to the parameters in $\bzeta_i$. 
It is convenient to note that 
(\ref{E0}) to (\ref{E3}) are analogous to that 
for the corresponding SMCFUSN mixture model, 
except that the scale and skewness matrices 
are now given by $\bSigma_i^* = \bB_i\bB_i^T+\bD_i$ 
and $\bDelta_i^* = \bB_i \bDelta_i$, respectively.  
In the case of a CFUSTFA model, for example, 
these are analogous to that for the FM-CFUST model 
which can be found in \citet{J164} 
and are also given in Appendix \ref{sec:appendA} for completeness. 
\newline

\noindent
The remaining conditional expectations can be derived 
by noting that the conditional distribution of $\bX_{ij}$ 
given $\bY_j$, $\bU_{ij}$, $w_{ij}$, and $Z_{ij}$ 
is a normal distribution. It can be shown that 
\begin{eqnarray}
\be_{3ij}^{(k)}	&=&	w_{ij}^{(k)} \bC_{i}^{(k)} 
		\bB_{i}^{(k)^T} \bD_{i}^{(k)^{-1}} 
		\left(\by_j - \bmu_{i}^{(k)}\right) 
		+ \bC_{i}^{(k)} \bDelta_{i}^{(k)} \be_{1ij}^{(k)}, 
		\label{X}\\
\be_{5ij}^{(k)}	&=& 
		\bC_{i}^{(k)} \bB_{i}^{(k)^T} \bD_{i}^{(k)^{-1}} 
		\left(\by_j - \bmu_{i}^{(k)}\right) \be_{1ij}^{(k)^T}  
		+ \bC_{i}^{(k)} \bDelta_{i}^{(k)} \be_{2ij}^{(k)},
		\label{Xt}\\
\be_{4ij}^{{(k)}}	&=& 
		\be_{3ij}^{(k)} \left(\by_j - \bmu_{i}^{(k)}\right)^T 
	  \bD_{i}^{(k)^{-1}} \bB_{i}^{(k)} \bC_{i}^{(k)^T}  
		+ \be_{5ij}^{(k)} \bDelta_{i}^{(k)^T}  
		\bC_{i}^{(k)^T} + \bC_{i}^{(k)}, 
		\label{Xs}
\end{eqnarray}
where $\bC_i^{(k)^{-1}} = \bB_i^{(k)^T} \bD_i^{(k)^{-1}} \bB_i^{(k)} + \bI_q$.

\subsection{CM-step}
\label{s:M}

The CM-steps are implemented by calculating 
the updated estimates of the parameters in $\bPsi$ 
by maximizing the $Q$-function obtained on the E-step. 
We proceed by updating the parameters in the order of 
$\pi_i$, $\bDelta_i$, $\bB_i$, $\bmu_i$, and $\bD_i$. 
More specifically, on the $(k+1)$th iteration of the ECM algorithm, 
the CM-steps are implemented as follows.
\newline

\noindent
CM-step 1: Compute the updated estimate of $\pi_i$ using
\begin{eqnarray}
\pi_i^{(k+1)}	&=&	\frac{1}{n} \sum_{j=1}^n z_{ij}^{(k)}. \nonumber
\end{eqnarray}
\newline

\noindent
CM-step 2: Compute the updated estimate of $\bDelta_i$ 
by maximising the $Q$-function over $\bDelta_i$, leading to
\begin{eqnarray}
\bDelta_i^{(k+1)}	&=&	
		\left[\sum_{j=1}^n z_{ij}^{(k)} \be_{5ij}^{(k)}\right]	
		\left[\sum_{j=1}^n z_{ij}^{(k)} \be_{2ij}^{(k)}\right]^{-1}. \nonumber
\end{eqnarray}
\newline

\noindent
CM-step 3: Fix $\bDelta_i = \bDelta_i^{(k)}$, then update $\bB_i$ using
\begin{eqnarray}
\bB_i^{(k+1)}	&=&	\left[\sum_{j=1}^n z_{ij}^{(k)} 
		\left(\by_j-\bmu_i^{(k+1)}\right) \be_{3ij}^{(k)^T}\right]	
		\left[\sum_{j=1}^n z_{ij}^{(k)} \be_{4ij}^{(k)}\right]^{-1}. \nonumber
\end{eqnarray}
\newline

\noindent
CM-step 4: Fix $\bDelta_i = \bDelta_i^{(k)}$ and $\bB_i = \bB_i^{(k+1)}$. 
The location vector $\bmu_i$ can be updated by
\begin{eqnarray}
\bmu_i^{(k+1)}	&=&	\frac{\sum_{j=1}^n z_{ij}^{(k)} w_{ij}^{(k)}\by_j 
		- \bB_i^{(k)}\bDelta_i^{(k)} \sum_{j=1}^n z_{ij}^{(k)} \be_{3ij}^{(k)}} 
		{\sum_{j=1}^n z_{ij}^{(k)} w_{ij}^{(k)}}. \nonumber
\end{eqnarray}
\newline

\noindent
CM-step 5: Fix $\bDelta_i = \bDelta_i^{(k)}$ and $\bmu_i = \bmu_i^{(k+1)}$. 
The updated estimate of $\bD_i$ can be obtained by 
maximizing the $Q$-function over $\bd_i$, 
the vector containing the diagonal elements of $\bD_i$. This leads to
\begin{eqnarray}
\bD_i^{(k+1)} &=& \mbox{diag}\left(\bd_i^{(k+1)}\right), \nonumber						
\end{eqnarray}
where
\begin{eqnarray}
\bd_i^{(k+1)}	 &=&	 \mbox{diag} \left\{
		\sum_{j=1}^n z_{ij}^{(k)} \left[ 
		\bB_i^{(k+1)} \be_{4ij}^{(k)} \bB_i^{(k+1)^T}  
		+ w_{ij}^{(k)} 
		\left(\by_j - \bmu_i^{(k+1)}\right) \left(\by_j-\bmu_i^{(k+1)}\right)^T
		\right.\right.\nonumber\\&&\left.\left. \hspace{-0.8cm} 
		- \bB_i^{(k+1)} \be_{3ij}^{(k)} \left(\by_j-\bmu_i^{(k+1)}\right)^T 
		- \left(\by_j-\bmu_i^{(k+1)}\right) \be_{3ij}^{(k)^T} \bB_i^{(k+1)^T} 
		\right] \Bigg\} \right.
		\left[\sum_{j=1}^n z_{ij}^{(k)}\right]^{-1} \hspace{-0.5cm}. \nonumber
\end{eqnarray} 
\newline

\noindent
CM-step 6: In the final CM-step, we compute 
the updated estimate of the parameters in $\bzeta_i$. 
Their expressions can be derived by maximizing the $Q$-function
with respect to $\bzeta_i$. 
In the case of the CFUSTFA model, for example, 
$\bzeta_i$ contains only $\nu_i$. 
An updated estimate of $\nu_i$ is obtained by 
solving for $\nu_i$ the following equation, 
\begin{eqnarray}
0 &=& \left(\sum_{i=1}^n z_{ij}^{(k)}\right) 
		\left[\log\left(\frac{\nu_i}{2}\right) 
		- \psi\left(\frac{\nu_i}{2}\right) + 1\right] \nonumber\\&&
		+ \sum_{j=1}^n z_{ij}^{(k)} \left[\psi\left(\frac{\nu_i^{(k)}+p}{2}\right)
		- \log\left(\frac{\nu_i^{(k)}+\eta_{ij}^{(k+1)}}{2}\right)
		- \left(\frac{\nu_i^{(k)}+p}{\nu_i^{(k)}+\eta_{ij}^{(k+1)}}\right) \right],
		\nonumber
\label{nu}
\end{eqnarray}
where
\begin{eqnarray}
\eta_{ij}^{(k+1)} &=& \left(\by_j - \bmu_i^{(k+1)}\right)^T 
\left(\bB_i^{(k+1)} \bOmega_i^{(k+1)} \bB_i^{(k+1)^T} + \bD_i^{(k+1)}\right)^{-1}  
\left(\by_j - \bmu_i^{(k+1)}\right), \nonumber\\
\bOmega_i^{(k+1)}	&=& \bI_q + \bDelta_i^{(k+1)} \bDelta_i^{(k+1)^T}, \nonumber
\end{eqnarray}
and where $\psi(\cdot)$ is the digamma function. 
\newline

\noindent
Given an initial value for the parameters in $\bPsi$, 
the ECM algorithm alternates between the E and CM-steps 
until a specified convergence criterion is met. 
These will be detailed in the next section. 
Upon convergence, the predicted component memberships 
can be obtained by applying 
the maximum \emph{a posteriori} (MAP) rule 
on the basis of the $z_{ij}^{(k)}$ \citep{B003}; 
that is, $\by_j$ is assigned to the component 
to which it has the highest posterior probability of belonging.    
In addition, factor scores can be useful for subsequent analysis, 
for example, to visualise the data in a lower-dimensional latent subspace. 
For the SMCFUSNFA model, the estimated factor scores 
can be easily obtained using (\ref{E4}), and are given by
\begin{eqnarray}
\hat{\bu}_{j} &=& \sum_{i=1}^g z_{ij}^{(k)} \be_{3ij}^{(k)}.
\label{scores}
\end{eqnarray}

\section{Implementation}
\label{s:impl}

\subsection{Starting Values}
\label{s:initial}

As the log likelihood function typically exhibits multiple local maxima, 
it is useful to try a variety of initial values 
using different starting strategies.    
An intuitive way to start the EM algorithm 
for the SMCFUSNFA model is to initialize the parameters 
according to the results of its nested model, for example, 
the corresponding restricted model with $r=1$, 
the symmetric version with $\bDelta=\bzero$, 
or the MFA model. 
In the latter case, as the MFA model 
does not account for skewness in the data, 
one may proceed by fitting a SMCFUSN distribution 
for the factor scores of each component of the MFA model 
and use the fitted parameters as the initial values.  

Alternatively, a convenient way to generate 
valid initial values for the SMCFUSNFA model 
is to start from an initial clustering of the data given by, 
for example, $k$-means, random partitions, 
or other clustering methods. 
We then proceed to fit a FA model to each cluster 
to obtain an initial estimate of $\bB_i$ and of $\bD_i$. 
Then, for the skewness matrix, its initial values 
can be obtained by fitting a SMCFUSN distribution 
(or any of its nested skew  distributions) 
to the factor scores of the FA model.

\subsection{Convergence}
\label{s:stop}

We monitor the convergence of the ECM algorithm 
using Aitken's acceleration criterion; 
see \citet[p.\ 137]{B004}.
More specifically, the algorithm is stopped 
when the absolute difference between 
the log likelihood value and the asymptotic log likelihood 
is less than $\epsilon = 10^{-6}$, that is, when
\begin{eqnarray}
\log L_\infty - \log L^{(k)} < \epsilon,
\label{converge}
\end{eqnarray}  
where $L^{(k)}$ denotes the likelihood value 
after the $k$th iteration of the EM algorithm 
and $L_{\infty}$ denotes the asymptotic estimate of the log likelihood. 

\subsection{Model selection}
\label{s:BIC}

In the ECM algorithm described above, 
the number of components $g$, 
the dimension of the latent factor subspace $q$, 
and the number of skewing variables $r$ are specified beforehand. 
In practice, these are typically unknown 
and need to be inferred from the data during model fitting. 
Usually, one proceeds by fitting the model 
for a range of values of $g$, $q$, and $r$ 
and to select an appropriate model from these candidates 
using some information criterion. 
The Bayesian information criterion (BIC) \citep{J056} 
is one of the more commonly used  criteria, and is defined as
\begin{eqnarray}
BIC &=& m \log(n) - 2 \log L(\hbPsi),
\label{BIC}
\end{eqnarray}
where $m$ is the number of free parameters, 
$n$ is the number of observations, 
and $L(\hat{\bPsi})$ is the maximised likelihood value. 
In addition, we consider also
 the integrated completed likelihood (ICL) criterion \citep{J625} 
to assist in choosing a suitable model. 
By construction, the ICL aims at finding 
the number of clusters in the data 
whereas the BIC is aimed at determining the optimal number of components. 
The former is more conservative 
and carries a heavier penalty for more complex models. 
The ICL is defined as
\begin{eqnarray}
ICL &=& BIC + 2 ENT,
\label{ICL}
\end{eqnarray}   
where $ENT = - \sum_{i=1}^g \sum_{j=1}^n 
\hat{z}_{ij} \log(\hat{z}_{ij})$ 
is the estimated partition mean entropy. 
Thus ICL can be considered as 
a entropy-penalized version of BIC that 
penalizes the overlap between mixture components. 
For both BIC and ICL, a smaller value of the criterion is preferred.

\section{\large Applications to real data}
\label{s:real}

In this section, we illustrate 
the application of the SMCFUSNFA model
in the particular case of the adoption of 
the CFUSTFA distribution.   
All analyses were performed in R \citep{S004}. 
For comparison, we consider also the GHSTFA, 
CGHSTFA, and GHFA models, 
and the nested models of CFUSTFA, namely, MSNFA and MFA. 
The GHFA model is implemented in 
the R package MixGHD \citep{S011}. 
The GHSTFA, CGHSTFA, and MSNFA models are implemented 
as in \citet{J618}, \citet{J619}, and \citet{J159b}, 
respectively. Initialization of model parameters 
and stopping rule are also implemented accordingly. 
Note that for the illustrations in this section, 
the number of components $g$ is assumed to be known for comparison purposes.    

For the datasets considered in this section, 
the true group labels are available and 
hence we can assess the clustering performance of these models. 
Here we consider the correct classification rate (CCR), 
the Adjusted Rand Index (ARI), 
and the Adjusted Mutual Information (AMI). 
The CCR ranges from 0 to 1.
It is calculated for all permutations of 
the cluster labels and the maximum CCR value 
across all permutations is reported. 
The ARI \citep{J123} is a variant of the Rand index \citep{J629} 
that corrects for chance so that 
it has a constant baseline equal to zero 
when the two clusterings are random and independent. 
An ARI of 1 indicating a perfect match to the `true' labels. 
The Adjusted Mutual Information (AMI)  \citep{J626}
is another popular measure 
used in the machine learning community. 
It is based on Shannon information theory 
whereas the ARI is based on pair-counting. 
Similar to ARI, the AMI is 
an adjusted version of the normalized mutual information 
that adjusts for chance. 
It has an expected value of zero for independent clusterings
and takes the maximum value of one  
when the two clusterings are in perfect agreement.

\subsection{The Hawks data}
\label{s:Hawks}

\begin{figure}
	\centering
		\includegraphics[width=1.00\textwidth]{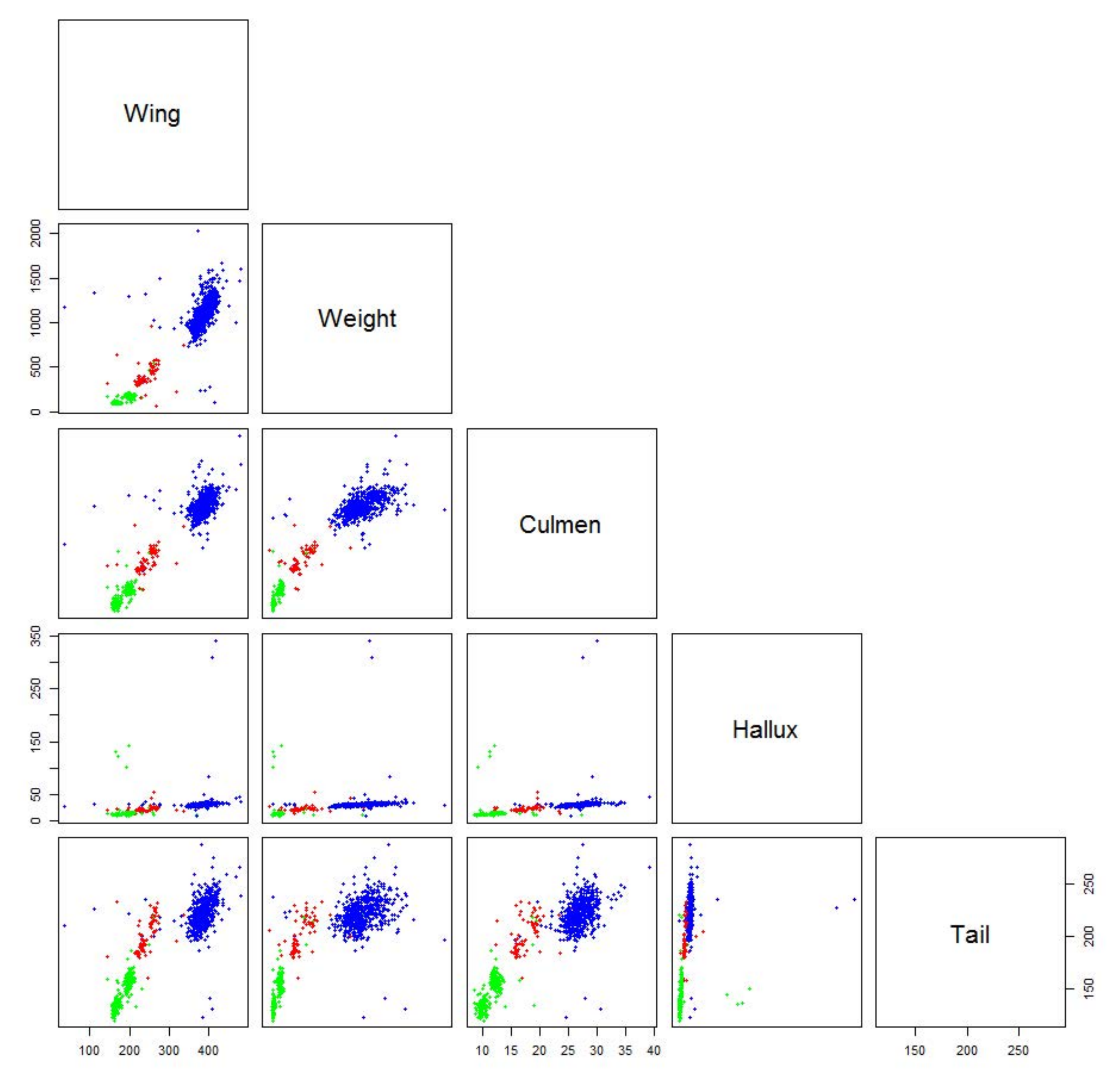}
	\caption{Bivariate scatterplot of the Hawks dataset. 
	Cooper's hawks are represented by red dots, 
	whereas red-tailed and sharp-shinned hawks 
	are represented by blue and green dots, respectively.  }
	\label{fig:Hawks}
\end{figure}

Our first illustration concerns a small dataset 
collected by researchers at the Cornell
College in Iowa. The data consist of 19 variables 
taken from hawks at Lake MacBride.
We consider here all the relevant continuous variables 
that have no missing observations. 
These are the length (mm) of primary wing feather, 
the weight (g) of the bird, 
the culmen length (mm), the hallux length (mm), 
and the tail length (mm). 
There are three species of hawks 
(Red-tailed, Sharp-shinned, and Copper's) 
in the data and a total of 891 samples. 
A summary of the data (Table \ref{tab:Hawks}) 
indicates that hallux length exhibit 
strong asymmetry and kurtosis, 
although it may not be clear from Figure \ref{fig:Hawks}. 
Mild skewness and kurtosis were also observed for the other variables. 
From Figure \ref{fig:Hawks}, there seems to be only mild overlap 
between the three clusters and hence we can expect 
the models to perform reasonably well in this dataset. 
A closer inspection of the dataset suggests that 
skewness appears to be concentrated in a single direction 
(or very close to it) and hence the skew factor models 
that are formulated using skew distributions 
with a single skewing variable 
should not be disadvantaged.

\begin{table*}[ht]
\centering
	\hspace*{-0.5cm} 
\begin{tabular}{|c|cccccc|}
  \hline
Variable & minimum & \makecell{sample \\ mean} & maximum & \makecell{sample \\ sd} 
	&  \makecell{sample \\ skewness} &  \makecell{sample \\ kurtosis} \\
  \hline
  Wing Length & 37.2 & 315.9 & 480.0 & 95.3 & -0.58 & 1.63 \\ 
  Weight & 56.0 & 771.6 & 2030.0 & 462.9 & -0.35 & 1.61 \\ 
  Culmen Length & 8.6 & 21.8 & 39.2 & 7.3 & -0.58 & 1.71 \\ 
  Hallux Length & 9.5 & 26.4 & 341.4 & 17.8 & 11.36 & 184.52 \\ 
  Tail Length & 119.0 & 198.9 & 288.0 & 36.8 & -0.70 & 2.16 \\ 
   \hline
\end{tabular}
\caption{An overview of the Hawks data.}
\label{tab:Hawks}
	\hspace*{-0.5cm} 
\end{table*}

\begin{figure*}
	\centering
		\includegraphics[width=1.00\textwidth]{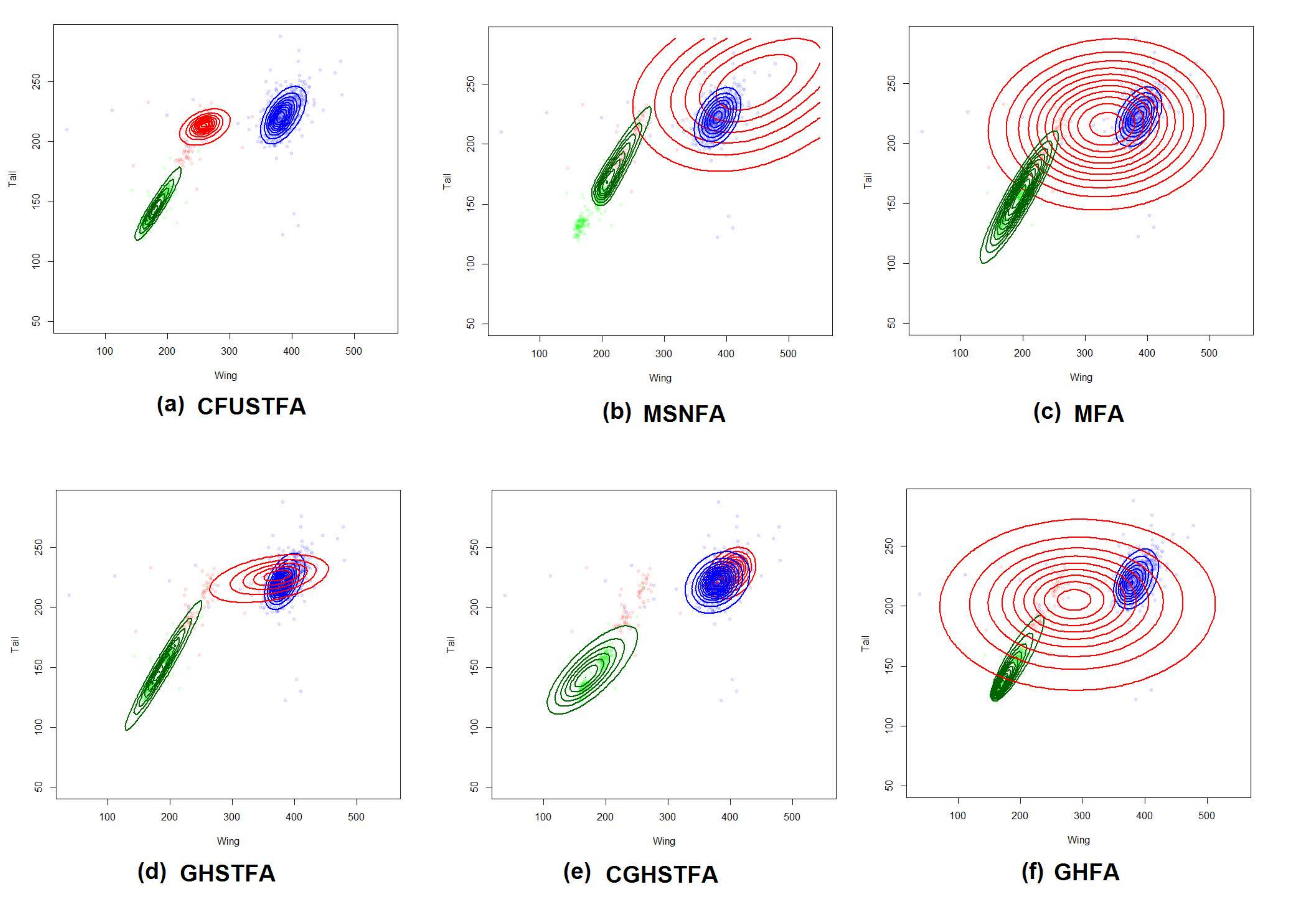}
	\caption{Contours of skew factor analyzers models fitted to the Hawks data. 
	Results are shown on the variables Tail and Wing.   }
	\label{fig:Hawks2}
\end{figure*}

\begin{figure*}
	\centering
		\includegraphics[width=1.00\textwidth]{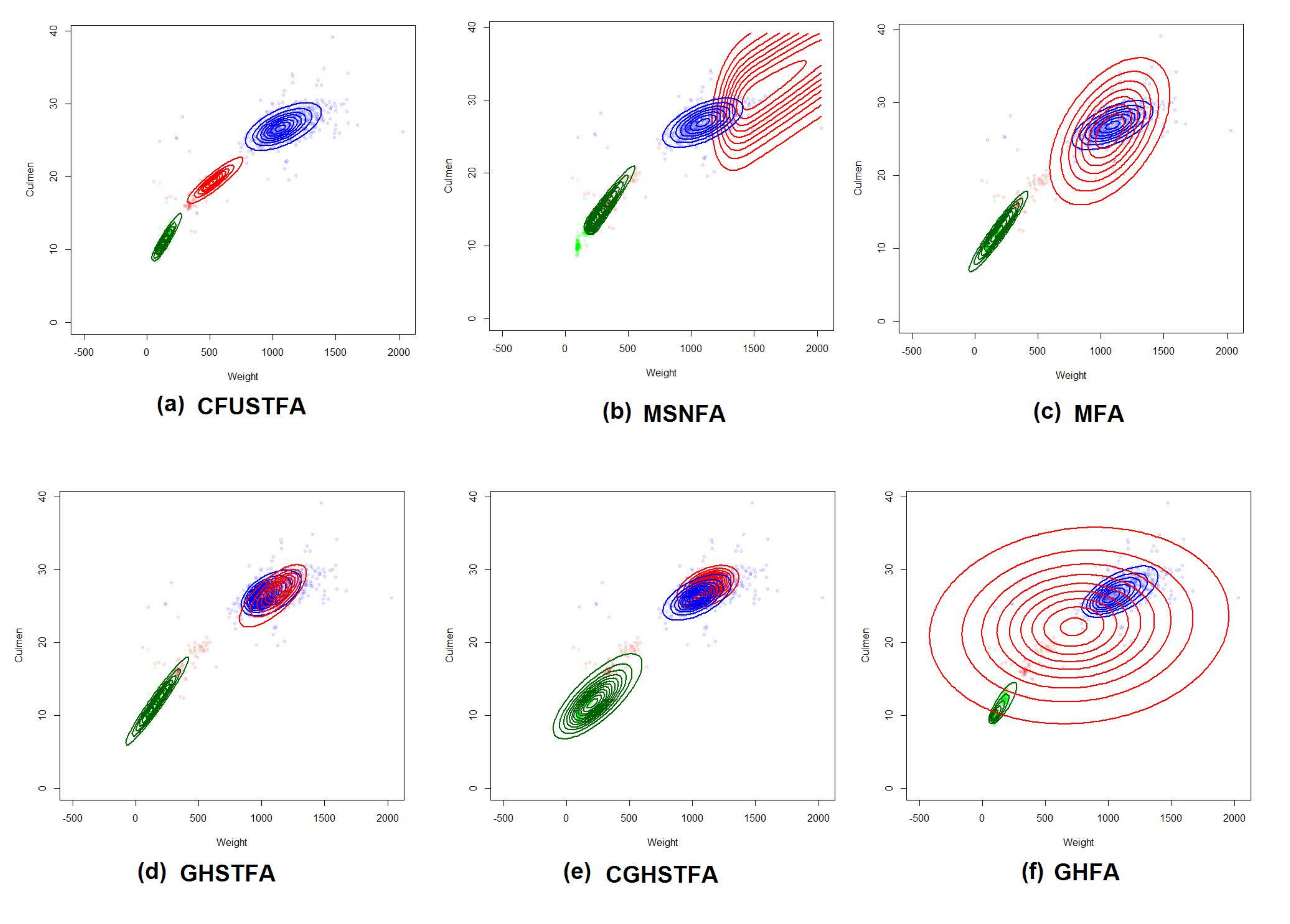}
	\caption{Contours of skew factor analyzers models fitted to the Hawks data. 
	Results are shown on the variables Culmen and Weight. }
	\label{fig:Hawks4}
\end{figure*}

\begin{table*}
	\centering
		\begin{tabular}{|c|c||c|c||c|c|c|}
			\hline
			Model		&	q	& BIC & ICL & CCR	&	ARI & AMI \\
			\hline
CFUSTFA & 2 & {32832} & {32872} & {0.9237} & {0.8783} & {0.7506} \\
  MSNFA & 1 & 33439 & 33469 & 0.88446 & 0.8045 & 0.6810 \\ 
  MFA & 1 & 33846 & 33878 & 0.8867 & 0.8069 & 0.7063 \\ 
  GHSTFA & 2 & 33457 & 33523 & 0.8945 & 0.8366 & 0.7189 \\ 
  CGHSTFA & 3 & 34630 & 34774 & 0.9136 & 0.8416 & 0.6826 \\ 
  GHFA & 2 & 33513 & 33486 & 0.8911 & 0.8267 & 0.7280 \\
			\hline
		\end{tabular}
	\caption{Performance of skew factor models on the Hawks data.}
	\label{tab:Hawks2}
\end{table*}

We fitted the CFUSTFA model with a range of values of $q$ and $r$ 
such that its number of free parameters is less than that of 
a finite mixture of CFUST distribution with 
the corresponding value of $r$. 
We also fitted the GHSTFA, the CGHSTFA, 
and the GHFA models and also the nested models MSNFA and MFA
with an appropriate range of values of $q$. 
Note that the range is dependent on the model. 
A summary of the preferred models by BIC 
is given in Table \ref{tab:Hawks2}. 
However, it is noted that better clustering results 
can be achieved for the models 
with slightly higher values of BIC 
than that reported in Table \ref{tab:Hawks2}. 

In this example for the Hawks data, 
the CFUSTFA model preferred by BIC had $r=1$ 
and thus corresponds to a MSTFA model. 
It can be seen from Table \ref{tab:Hawks2} that 
the CFUSTFA obtained the highest CCR, ARI, and AMI.  
It is also preferred over the other models according to BIC and ICL.
The MSNFA model is ranked second by BIC and ICL. 
The next two preferred models according to BIC and ICL 
are the GHSTFA and GHFA models, where the former is 
preferred over the latter by BIC 
and the opposite is preferred by ICL. 
According to Table \ref{tab:Hawks2}, 
the GHSTFA model obtained slightly better clustering performance
than the GHFA model according to CCR and ARI, 
although the AMI ranked the clustering results obtained by 
the GHFA model slightly more preferable than that of the GHSTFA model.    
We can also observe from Table \ref{tab:Hawks2}
that the GHSTFA and CGHSTFA models 
have very similar performance. 
The CGHSTFA model gave slightly better clustering results 
in terms of the CCR and ARI than the GHSTFA model, 
but the latter model is not preferred to the CGHSTFA model 
in terms of BIC and ICL. 
This can be partly observed from Figures \ref{fig:Hawks2}(d) and 
\ref{fig:Hawks2}(e),  
where the CGHSTFA model appears to provide a closer fit to the data 
than the GHSTFA model.  
In particular, the location and shape of the component 
shown in red in Figures \ref{fig:Hawks2} and \ref{fig:Hawks4} 
are quite different.  

On comparing the contours of the models in 
Figures \ref{fig:Hawks2} and \ref{fig:Hawks4}, 
it can be observed that all of the considered models
except the CFUSTFA model seem to have difficulty 
separating the two upper clusters (shown in red and blue).             
Overall, the visual impression from 
Figures \ref{fig:Hawks2} and \ref{fig:Hawks4} 
supports the preference by BIC and ICL 
which suggests that the CFUSTFA model provides 
a better fit relative to the other models considered in this dataset.

\subsection{The melanoma data}
\label{s:melan}

\begin{figure*}
	\centering
		\includegraphics[width=0.8\textwidth]{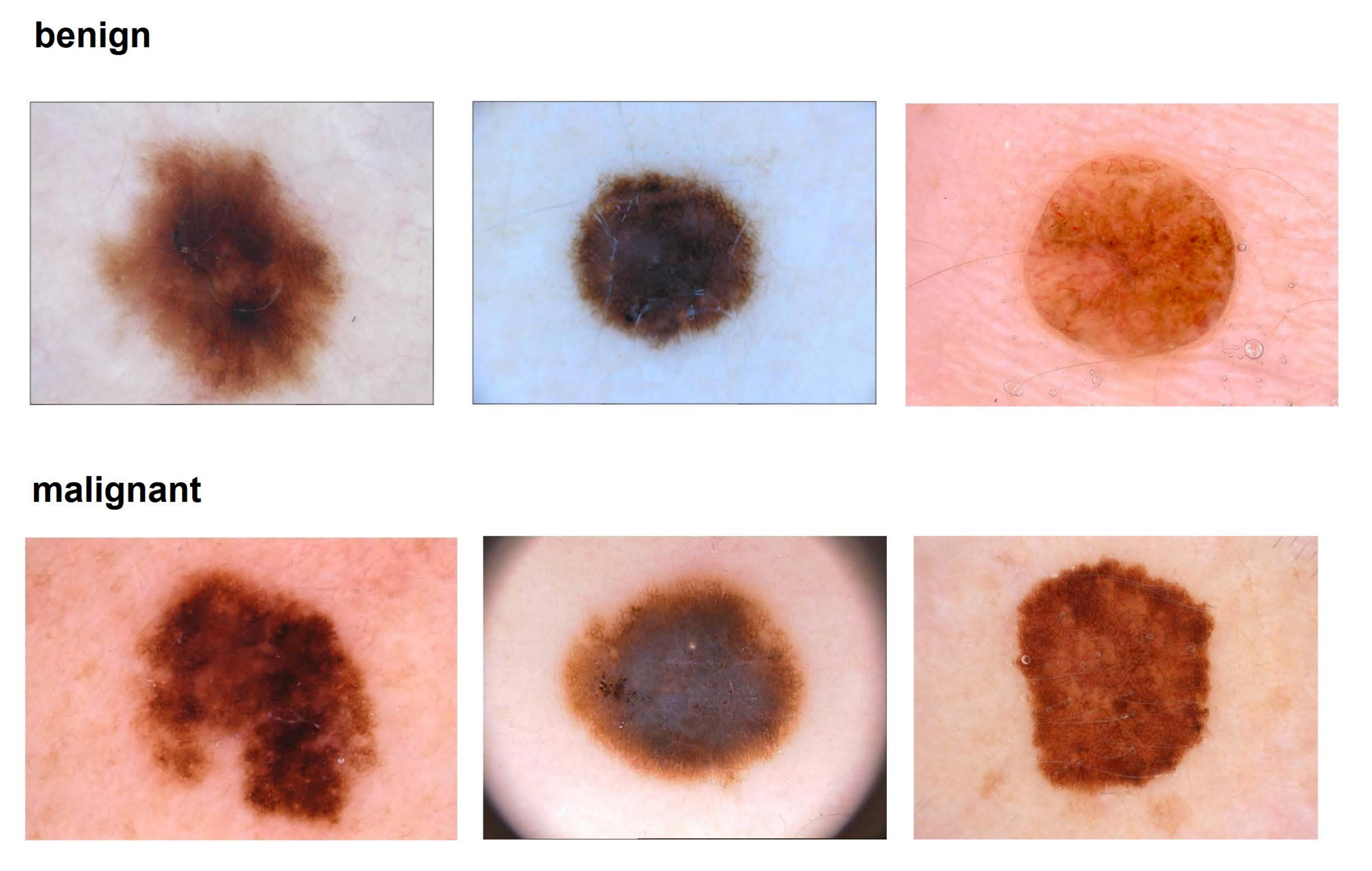}
	\caption{Example of benign and malignant images from the ISIC Archive.}
	\label{fig:melanoma}
\end{figure*}

We consider an application of mixtures of factor analyzers 
to the discrimination between benign and malignant melanoma 
from clinical and dermoscopic skin images. 
Images can be obtained from public databases such as ISIC \citep{J613}; 
some examples are shown in Figure \ref{fig:melanoma}. 
Commonly used features for medical image processing 
were extracted from these images. 
These include some of those suggested by \citet{J614}, 
such as eccentricity, equivalent diameter, perimeter, and solidity. 
For this illustration, 149 cases of benign lesions 
and 149 cases of malignant lesions were included -- 
a total 298 images to be analyzed. 
We considered the fitting of the CFUSTFA, MSNFA, MFA, 
GHSTFA, CGHSTFA, and GHFA models to the data with $g=2$. 
The models were applied with $q$ varying from $1$ and $10$ 
(the maximum value of $q$ is dependent on the model). 
The best performing results are reported in Table \ref{tab:melanoma}. 
As can be observed from these results, 
this is a difficult dataset for clustering. 
The best clustering results are obtained by the CFUSTFA model  
which has an ARI of 0.57 and a CCR of 0.88. 
The next best performing model according to CCR and ARI 
is the GHFA model. 
However, its CCR is considerably lower (approximately 20\% less) 
than that for the CFUSTFA model
and its ARI is very low (ARI=0.11). 
The remaining models have similar performance to the GHSTFA model, 
which is the next best performing model according to CCR and ARI, 
as can be observed the results in Table \ref{tab:melanoma}.  
A cross-tabulation of the clustering results of 
the best performing models is given in Table \ref{tab:melanoma2}. 
With the CFUSTFA model, there are 36 misclassified observations, 
whereas with the GHFA and GHSTFA models, 
there are 94 and 108 misclassified observations, respectively. 
In addition, it is of interest to note that 
the CFUSTFA model has only $q=3$ factors, 
whereas 
the other five models require more factors for this dataset (ranging from 7 to 10).        

\begin{table*}
	\centering
		\begin{tabular}{|c|c||c|c||c|c|c|c|}
			\hline
			Model		&	q	& BIC & ICL & CCR	&	ARI & AMI\\
			\hline
 CFUSTFA & 3 & 24193 & 24201 & {0.8792} & {0.5738} & {0.5007} \\ 
  MSNFA & 10 & 22115 & 22125 & 0.6208 & 0.0560 & 0.0571 \\ 
  MFA & 10 & 22253 & 22271 & 0.6376 & 0.0733 & 0.0718 \\ 
  GHSTFA & 10 & {22092} & {22104} & 0.6376 & 0.0736 & 0.0827 \\ 
  CGHSTFA & 8 & 25282 & 25331 & 0.6309 & 0.0670 & 0.1172 \\ 
  GHFA & 7 & 28486 & 28465 & 0.6846 & 0.1336 & 0.1118 \\ 
	\hline
		\end{tabular}
	\caption{Performance of skew factor models for the melanoma data.}
	\label{tab:melanoma}
\end{table*}

\begin{table}
	\centering
		\begin{tabular}{|c||cc||cc||cc|}
			\hline
					&	\multicolumn{2}{c||}{CFUSTFA} &	\multicolumn{2}{c||}{GHSTFA} 
					& \multicolumn{2}{c|}{GHFA}\\ \hline\hline
					benign		&	144 & 5	 & 137 & 12 & 126 & 23  \\
					malignant	&	31 & 118 & 96 & 53  &  71 & 78 \\
			\hline
		\end{tabular}
	\caption{Cross-tabulation of clustering results of 
	the CFUSTFA, GHSTFA, and GHFA models against 
	the true group labels of the melanoma data.}
	\label{tab:melanoma2}
\end{table}

\section{\large Conclusions}
\label{s:concl}

This paper presents a novel generalization of 
the mixture of factor analyzers model based on 
a general skew distributional form 
that defines the class of SMCFUSN distributions. 
The proposed model provides a powerful tool 
for the flexible modelling of data exhibiting non-normal features 
including multimodality, skewness, and heavy-tailedness. 
For illustration, the focus has been on 
its special case, 
namely, the CFUST distribution. An ECM algorithm is 
derived for the mixture of SMCFUSN factor analzyers. 
Implementation issues such as strategies for 
generating starting values, the choice of convergence assessment, 
and model selection tools are also discussed. 
This class of mixture of skew factor analyzers 
formally embeds most of the existing skew factor analyzers, 
including the MSNFA, MSTFA, and CFUSSHFA models by \citet{J159b}, 
\citet{J612}, and \citet{J615}, respectively. 
An investigation of various existing mixtures of 
skew factor analyzers is presented, 
outlining the links and differences between them.  
Unlike existing models that are based on 
restricted skew distributions, 
the proposed SMCFUSNFA model has the capability of 
modelling multiple arbitrary directions of skewness at the same time.  
The usefulness of the SMCFUSNFA model is illustrated 
using the CFUSTFA model on some real datasets 
and its effectiveness over competing models 
is demonstrated in terms of various performance assessment measures.

\appendix
\section{Expressions for the E-step of the ECM algorithm for the CFUSTFA model}
\label{sec:appendA}

For the CFUSTFA model, the E-step of the ECM algorithms 
involves four conditional expressions that are 
analogous to the case of mixtures of CFUST distributions. 
Technical details can be found in \citet{J164}. 
The expressions for (\ref{E0}) to (\ref{E3}) 
are similar to that for (12), (13), (15), and (16), 
respectively, in \citet{J164}. 
However, the scale matrices and skewness matrices 
in our case are given by 
$\bSigma_i^{*^{(k)}} = \bB_i^{(k)}\bB_i^{(k)}+\bD_i^{(k)}$ 
and $\bDelta_i^{*^{(k)}} = \bB_i^{(k)} \bDelta_i^{(k)}$ 
$(i=1, \ldots, g)$, respectively. 
Thus, the expressions for the conditional expectations 
(\ref{E0}) to (\ref{E3}) are given by
\begin{eqnarray}
z_{ij}^{(k)} &=&	
			\frac{\pi_i^{(k)} f_{\mbox{\tiny{CFUST}}_{p,r}}
			(\by_j; \bmu_i^{(k)}, \bSigma_i^{*^{(k)}}, \bDelta_i^{*^{(k)}}, \nu_i^{(k)})}
			{\sum_{i=1}^g \pi_i^{(k)} f_{\mbox{\tiny{CFUST}}_{p,r}}
			(\by_j; \bmu_i^{(k)}, \bSigma_i^{*^{(k)}}, \bDelta_i^{*^{(k)}}, \nu_i^{(k)})}, 
			\label{e0}\\
w_{ij}^{(k)}	&=& 		
		  \left(\frac{\nu_i^{(k)} + p}{\nu_i^{(k)} + d_{ij}^{(k)}}\right)  
			\frac{T_r\left(\bc_{ij}^{(k)} \sqrt{\frac{\nu_i^{(k)}+p+2}{\nu_i+d_{ij}^{(k)}}}; 
				\bzero, \bLambda_i^{(k)}, \nu_i^{(k)}+p+2\right)}
				{T_r\left(\bc_{ij}^{(k)} \sqrt{\frac{\nu_i^{(k)}+p}{\nu_i+d_{ij}^{(k)}}}; 
				\bzero, \bLambda_i^{(k)}, \nu_i^{(k)}+p\right)}, 
				\label{e1}\\
\be_{1ij}^{(k)}	&=&	
			w_{ij}^{(k)} E\left[ \ba_{ij}^{(k)}\right], 	\label{e4}\\
\be_{2ij}^{{(k)}} &=&	
			w_{ij}^{(k)} E\left[\ba_{ij}^{(k)} \ba_{ij}^{(k)^T} \right],	\label{e5}
\end{eqnarray} 
where
\begin{eqnarray}
d_{ij}^{(k)}	&=&	(\by_j - \bmu_i^{(k)})^T \bOmega_i^{(k)^{-1}} (\by_i - \bmu_i^{(k)}), 
		\nonumber\\
\bc_{ij}^{(k)}	&=& \bDelta_i^{*^{(k)^T}} \bOmega_i^{(k)^{-1}} (\by_j-\bmu_i^{(k)}), 
		\nonumber\\
\bLambda_i^{(k)}	&=& \bI_r - \bDelta_i^{*^{(k)^T}} \bOmega_i^{(k)^{-1}} \bDelta_i^{(k)}, 
		\nonumber\\
\bOmega_i^{(k)}	&=& \bSigma_i^{*^{(k)}} + \bDelta_i^{*^{(k)}} \bDelta_i^{*^{(k)^T}}, 
		\nonumber
\end{eqnarray}
and where $\ba_{ij}^{(k)}$ is a $r$-variate 
truncated $t$-random variable given by
\begin{eqnarray}
\ba_{ij}^{(k)}	&\sim& Tt_r\left(\bc_{ij}^{(k)}, 
		\left(\frac{\nu_i^{(k)} + d_{ij}^{(k)}}{\nu_i^{(k)}+p+2}\right) 
		\bLambda_i^{(k)}, \nu_i^{(k)}+p+2; \mathbb{R}^+\right). \nonumber
\end{eqnarray}
The last term in expressions (\ref{e4}) and (\ref{e5}) 
correspond to the first and second moment of $\ba_{ij}^{(k)}$ 
and can be evaluated using formulae described in, 
for example, \citet{W065}, \citet{J067}, 
and in the appendix of \citet{J103}.



\begin{thebibliography}{53}
\providecommand{\natexlab}[1]{#1}
\providecommand{\url}[1]{\texttt{#1}}
\providecommand{\urlprefix}{URL }

\bibitem[{Arellano-Valle and Azzalini(2006)}]{J017}
Arellano-Valle, R.B. and Azzalini, A. (2006).
\newblock On the unification of families of skew-normal distributions.
\newblock \emph{Scandinavian Journal of Statistics} \textbf{33}, 561--574.

\bibitem[{Arellano-Valle and Genton(2005)}]{J008}
Arellano-Valle, R.B. and Genton, M.G. (2005).
\newblock On fundamental skew distributions.
\newblock \emph{Journal of Multivariate Analysis} \textbf{96}, 93--116.

\bibitem[{Azzalini and Capitanio(2014)}]{B025}
Azzalini, A. and Capitanio, A. (2014).
\newblock \emph{The Skew-Normal and Related Families}.
\newblock Cambridge: Cambridge University Press.

\bibitem[{Azzalini and {\uppercase{D}alla Valle}(1996)}]{J001}
Azzalini, A. and {\uppercase{D}alla Valle}, A. (1996).
\newblock The multivariate skew-normal distribution.
\newblock \emph{Biometrika} \textbf{83}, 715--726.

\bibitem[{Biernacki et~al.(2000)Biernacki, Celeux, and Govaert}]{J625}
Biernacki, C., Celeux, G., and Govaert, G. (2000).
\newblock Assessing a mixture model for clustering with the integrated
  completed likelihood.
\newblock \emph{IEEE Transactions on Pattern Analysis and Machine Intelligence}
  \textbf{22}, 719--725.

\bibitem[{Browne and McNicholas(2015)}]{J401}
Browne, R.P. and McNicholas, P.D. (2015).
\newblock A mixture of generalized hyperbolic distributions.
\newblock \emph{The Canadian Journal of Statistics} \textbf{43}, 176--198.

\bibitem[{Cabral et~al.(2012)Cabral, Lachos, and Prates}]{J066}
Cabral, C.R.B., Lachos, V.H., and Prates, M.O. (2012).
\newblock Multivariate mixture modeling using skew-normal independent
  distributions.
\newblock \emph{Computational Statistics and Data Analysis} \textbf{56},
  126--142.

\bibitem[{Codella et~al.(2017)Codella, Gutman, Celebi, Helba, Marchetti, Dusza,
  Kalloo, Liopyris, Mishra, Kittler, and Halpern}]{J613}
Codella, N., Gutman, D., Celebi, M.E., Helba, B., Marchetti, M.A., Dusza, S.,
  Kalloo, A., Liopyris, K., Mishra, N., Kittler, H., and Halpern, A. (2017).
\newblock Skin lesion analysis toward melanoma detection: A challenge at the
  2017 {International Symposium on Biomedical Imaging (ISBI)}, hosted by the
  {International Skin Imaging Collaboration (ISIC)}.
\newblock \emph{arXiv:1710.05006} .

\bibitem[{Dempster et~al.(1977)Dempster, Laird, and Rubin}]{J034}
Dempster, A.P., Laird, N.M., and Rubin, D.B. (1977).
\newblock Maximum likelihood from incomplete data via the {EM} algorithm.
\newblock \emph{Journal of Royal Statistical Society B} \textbf{39}, 1--38.

\bibitem[{Ferris et~al.(2015)Ferris, Harkes, Gilbert, Winger, Golubets, Akilov,
  and Satyanarayanan}]{J614}
Ferris, L.K., Harkes, J.A., Gilbert, B., Winger, D.G., Golubets, K., Akilov,
  O., and Satyanarayanan, M. (2015).
\newblock Computer-aided classification of melanocytic lesions using
  dermoscopic images.
\newblock \emph{Journal of the American Academy of Dermatology} \textbf{73},
  769--776.

\bibitem[{Genton(2004)}]{B001}
Genton, M.G. (Ed.). (2004).
\newblock \emph{Skew-Elliptical Distributions and Their Applications: A Journey
  Beyond Normality}.
\newblock Boca Raton, Florida: Chapman \& Hall, CRC.

\bibitem[{Ghahramani and Hinton(1997)}]{J617}
Ghahramani, Z. and Hinton, G. (1997).
\newblock The {EM} algorithm for factor analyzers.
\newblock \emph{Technical Report No. CRG-TR-96-1} The University of Toronto:
  Toronto.

\bibitem[{Ho et~al.(2012)Ho, Lin, Chen, and Wang}]{J067}
Ho, H.J., Lin, T.I., Chen, H.Y., and Wang, W.L. (2012).
\newblock Some results on the truncated multivariate $t$ distribution.
\newblock \emph{Journal of Statistical Planning and Inference} \textbf{142},
  25--40.

\bibitem[{Hubert and Arabie(1985)}]{J123}
Hubert, L. and Arabie, P. (1985).
\newblock Comparing partitions.
\newblock \emph{Journal of Classification} \textbf{2}, 193--218.

\bibitem[{Karlis and Santourian(2009)}]{J375}
Karlis, D. and Santourian, A. (2009).
\newblock Model-based clustering with non-elliptically contoured distributions.
\newblock \emph{Statistics and Computing} \textbf{19}, 73--83.

\bibitem[{Kim et~al.(2016)Kim, Maadooliat, Arellano-Valle, and Genton}]{J624}
Kim, H.M., Maadooliat, M., Arellano-Valle, R.B., and Genton, M.G. (2016).
\newblock Skewed factor models using selection mechanisms.
\newblock \emph{Journal of Multivariate Analysis} \textbf{145}, 162--177.

\bibitem[{Kim(2016)}]{J451}
Kim, S.G. (2016).
\newblock An approximate fitting for mixture of multivariate skew normal
  distribution via {EM} algorithm.
\newblock \emph{Korean Journal of Applied Statistics} \textbf{29}, 513--523.

\bibitem[{Lee and McLachlan(2014)}]{J103}
Lee, S. and McLachlan, G.J. (2014).
\newblock Finite mixtures of multivariate skew $t$-distributions: Some recent
  and new results.
\newblock \emph{Statistics and Computing} \textbf{24}, 181--202.

\bibitem[{Lee and McLachlan(2013)}]{J105}
Lee, S.X. and McLachlan, G.J. (2013).
\newblock On mixtures of skew-normal and skew $t$-distributions.
\newblock \emph{Advances in Data Analysis and Classification} \textbf{7},
  241--266.

\bibitem[{Lee and McLachlan(2016)}]{J164}
Lee, S.X. and McLachlan, G.J. (2016).
\newblock Finite mixtures of canonical fundamental skew $t$-distributions: The
  unification of the restricted and unrestricted skew $t$-mixture models.
\newblock \emph{Statistics and Computing} \textbf{26}, 573--589.

\bibitem[{Lichman(2013)}]{UCI}
Lichman, M. (2013).
\newblock {UCI} machine learning repository.
\newblock \urlprefix\url{http://archive.ics.uci.edu/ml}.

\bibitem[{Lin(2009)}]{J621}
Lin, T.I. (2009).
\newblock Maximum likelihood estimation for multivariate skew normal mixture
  models.
\newblock \emph{Journal of Multivariate Analysis} \textbf{100}, 257--265.

\bibitem[{Lin(2010)}]{J027}
Lin, T.I. (2010).
\newblock Robust mixture modeling using multivariate skew-$t$ distribution.
\newblock \emph{Statistics and Computing} \textbf{20}, 343--356.

\bibitem[{Lin et~al.(2016)Lin, McLachlan, and Lee}]{J159b}
Lin, T.I., McLachlan, G.J., and Lee, S.X. (2016).
\newblock Extending mixtures of factor models using the restricted multivariate
  skew-normal distribution.
\newblock \emph{Journal of Multivariate Analysis} \textbf{143}, 398--413.

\bibitem[{Lin et~al.(2018)Lin, Wang, McLachlan, and Lee}]{J612}
Lin, T.I., Wang, W.L., McLachlan, G.J., and Lee, S.X. (2018).
\newblock Robust mixtures of factor analysis models using the restricted
  multivariate skew-$t$ distribution.
\newblock \emph{Statistical Modelling} \textbf{18}, 50--72.

\bibitem[{Lin et~al.(2015)Lin, Wu, McLachlan, and Lee}]{J160b}
Lin, T.I., Wu, P.H., McLachlan, G.J., and Lee, S.X. (2015).
\newblock A robust factor analysis model using the restricted skew
  $t$-distribution.
\newblock \emph{TEST} \textbf{24}, 510--531.

\bibitem[{Maleki et~al.(2018)Maleki, Wraith, and Arellano-Valle}]{J631}
Maleki, M., Wraith, D., and Arellano-Valle, R.B. (2018).
\newblock Robust finite mixture modeling of multivariate unrestricted
  skew-normal generalized hyperbolic distributions.
\newblock \emph{Statistics and Computing} .

\bibitem[{Maruotti et~al.(2017)Maruotti, Bulla, Lagona, Picone, and
  Martella}]{J610}
Maruotti, A., Bulla, J., Lagona, F., Picone, M., and Martella, F. (2017).
\newblock Dynamic mixtures of factor analyzers to characterize multivariate air
  pollutant exposures.
\newblock \emph{Annals of Applied Statistics} \textbf{3}, 1617--1648.

\bibitem[{McLachlan et~al.(2007)McLachlan, Bean, and Jones}]{J622}
McLachlan, G.J., Bean, R.W., and Jones, B.T. (2007).
\newblock Extension of the mixture of factor analyzers model to incorporate the
  multivariate $t$-distribution.
\newblock \emph{Computational Statistics and Data Analysis} \textbf{51},
  5327--–5338.

\bibitem[{McLachlan and Krishnan(2008)}]{B004}
McLachlan, G.J. and Krishnan, T. (2008).
\newblock  (Second Edition).
\newblock \emph{The {EM} Algorithm and Extensions}.
\newblock Hoboken, New Jersey: Wiley.

\bibitem[{McLachlan and Lee(2016)}]{J360b}
McLachlan, G.J. and Lee, S.X. (2016).
\newblock Comment on ``{On} nomenclature for, and the relative merits of, two
  formulations of skew distributions'' by {A. Azzalini, R. Browne, M. Genton,
  and P. McNicholas}.
\newblock \emph{Statistics and Probability Letters} \textbf{116}, 1--5.

\bibitem[{McLachlan and Peel(2000)}]{B003}
McLachlan, G.J. and Peel, D. (2000).
\newblock \emph{Finite Mixture Models}.
\newblock New York: Wiley.

\bibitem[{McLachlan et~al.(2003)McLachlan, Peel, and Bean}]{J608}
McLachlan, G.J., Peel, D., and Bean, R.W. (2003).
\newblock Modelling high-dimensional data by mixtures of factor analyzers.
\newblock \emph{Computational Statistics and Data Analysis} \textbf{41},
  379--388.

\bibitem[{Meng and Rubin(1993)}]{J630}
Meng, X. and Rubin, D. (1993).
\newblock Maximum likelihood estimation via the {ECM} algorithm: a general
  framework.
\newblock \emph{Biometrika} \textbf{80}, 267--278.

\bibitem[{Montanari and Viroli(2010)}]{J611}
Montanari, A. and Viroli, C. (2010).
\newblock A skew-normal factor model for the analysis of student satisfaction
  towards university courses.
\newblock \emph{Journal of Applied Statistics} \textbf{37}, 463--487.

\bibitem[{Murray et~al.(2014{\natexlab{a}})Murray, Browne, and
  McNicholas}]{J618}
Murray, P., Browne, R., and McNicholas, P. (2014{\natexlab{a}}).
\newblock Mixtures of skew-$t$ factor analyzers.
\newblock \emph{Computational Statistics and Data Analysis} \textbf{77},
  326--335.

\bibitem[{Murray et~al.(2014{\natexlab{b}})Murray, McNicholas, and
  Browne}]{J619}
Murray, P., McNicholas, P., and Browne, R. (2014{\natexlab{b}}).
\newblock Mixtures of common skew-$t$ factor analyzers.
\newblock \emph{Stat} \textbf{3}, 68--82.

\bibitem[{Murray et~al.(2017{\natexlab{a}})Murray, Browne, and
  McNicholas}]{J616}
Murray, P.M., Browne, R.P., and McNicholas, P.D. (2017{\natexlab{a}}).
\newblock Hidden truncation hyperbolic distributions, finite mixtures thereof,
  and their application for clustering.
\newblock \emph{Journal of Multivariate Analysis} \textbf{161}, 141--156.

\bibitem[{Murray et~al.(2017{\natexlab{b}})Murray, Browne, and
  McNicholas}]{J623}
Murray, P.M., Browne, R.P., and McNicholas, P.D. (2017{\natexlab{b}}).
\newblock A mixture of {SDB} skew-$t$ factor analyzers.
\newblock \emph{Econometrics and Statistics} \textbf{3}, 160--168.

\bibitem[{Murray et~al.(2017{\natexlab{c}})Murray, Browne, and
  McNicholas}]{J615}
Murray, P.M., Browne, R.P., and McNicholas, P.D. (2017{\natexlab{c}}).
\newblock Mixtures of hidden truncation hyperbolic factor analyzers.
\newblock \emph{arXiv:1711.01504} .

\bibitem[{O'Hagan(1976)}]{W065}
O'Hagan, A. (1976).
\newblock Moments of the truncated multivariate-$t$ distribution.
\newblock \url{http://www.tonyohagan.co.uk/academic /pdf/trunc\_multi\_t.PDF}.

\bibitem[{Pyne et~al.(2009)Pyne, Hu, Wang, Rossin, Lin, Maier, Baecher-Allan,
  McLachlan, Tamayo, Hafler, De~Jager, and Mesirow}]{J004}
Pyne, S., Hu, X., Wang, K., Rossin, E., Lin, T.I., Maier, L.M., Baecher-Allan,
  C., McLachlan, G.J., Tamayo, P., Hafler, D.A., De~Jager, P.L., and Mesirow,
  J.P. (2009).
\newblock Automated high-dimensional flow cytometric data analysis.
\newblock \emph{Proceedings of the National Academy of Sciences USA}
  \textbf{106}, 8519--8524.

\bibitem[{{R Core Team}(2016)}]{S004}
{R Core Team} (2016).
\newblock \emph{{R}: A Language and Environment for Statistical Computing}.
\newblock \urlprefix\url{http://www.R-project.org/}.
\newblock {R} Foundation for Statistical Computing, Vienna, Austria. {ISBN
  3-900051-07-0}.

\bibitem[{Rand(1971)}]{J629}
Rand, W.M. (1971).
\newblock Objective criteria for the evaluation of clustering methods.
\newblock \emph{Journal of the American Statistical Association} \textbf{66},
  846--850.

\bibitem[{Sahu et~al.(2003)Sahu, Dey, and Branco}]{J002}
Sahu, S.K., Dey, D.K., and Branco, M.D. (2003).
\newblock A new class of multivariate skew distributions with applications to
  {B}ayesian regression models.
\newblock \emph{The Canadian Journal of Statistics} \textbf{31}, 129--150.

\bibitem[{Schwarz(1978)}]{J056}
Schwarz, G. (1978).
\newblock Estimating the dimension of a model.
\newblock \emph{Annals of Statistics} \textbf{6}, 461--464.

\bibitem[{Seshadri(1997)}]{B104}
Seshadri, V. (1997).
\newblock Halphen's laws.
\newblock In \emph{Encyclopedia of Statistical Sciences},  S.~Kotz, C.~B. Read,
  and D.~L. Banks  (Eds.). New York: Wiley, pp.\ 302--306.

\bibitem[{Tortora et~al.(2015)Tortora, Browne, Franczak, and McNicholas}]{S011}
Tortora, C., Browne, R.P., Franczak, B.C., and McNicholas, P.D. (2015).
\newblock \emph{MixGHD: Model Based Clustering, Classification and Discriminant
  Analysis Using the Mixture of Generalized Hyperbolic Distributions.}
\newblock \urlprefix\url{http://cran.r-project.org/web/packages/MixGHD}.
\newblock R package version 1.7.

\bibitem[{Tortora et~al.(2016)Tortora, McNicholas, and Browne}]{J620}
Tortora, C., McNicholas, P., and Browne, R. (2016).
\newblock A mixture of generalized hyperbolic factor analyzers.
\newblock \emph{Advances in Data Analysis and Classification} \textbf{10},
  423--440.

\bibitem[{Vinh et~al.(2010)Vinh, Epps, and Bailey}]{J626}
Vinh, N.X., Epps, J., and Bailey, J. (2010).
\newblock Information theoretic measures for clusterings comparison: Variants,
  properties, normalization and correction for chance.
\newblock \emph{Journal of Machine Learning Research} \textbf{11}, 2227--2240.

\bibitem[{Wall et~al.(2012)Wall, Guo, and Amemiya}]{J609}
Wall, M.M., Guo, J., and Amemiya, Y. (2012).
\newblock Mixture factor analysis for approximating a non-normally distributed
  continuous latent factor with continuous and dichotomous observed variables.
\newblock \emph{Multivariate Behavioral Research} \textbf{47}, 276--313.

\bibitem[{Yamamoto et~al.(2005)Yamamoto, Nankaku, Miyajima, Tokuda, and
  Kitamura}]{J606}
Yamamoto, H., Nankaku, Y., Miyajima, C., Tokuda, K., and Kitamura, T. (2005).
\newblock Parameter sharing in mixture of factor analyzers for speaker
  identification.
\newblock \emph{IEICE Transactions on Information and Systems} \textbf{88},
  418--424.

\bibitem[{Zhoe and Mobasher(2006)}]{J607}
Zhoe, Y.K. and Mobasher, B. (2006).
\newblock Web user segmentation based on a mixture of factor analyzers.
\newblock \emph{Lecture Notes in Computer Science} \textbf{4082}, 11--20.

\end{thebibliography}
\end{document}